\documentclass[12pt]{article}
\usepackage{amsmath,amssymb}
\usepackage{graphicx,color}
\usepackage[linkcolor=blue,colorlinks=true]{hyperref}
\usepackage{cite,epsf}
\usepackage[bf]{caption}
\usepackage{subfig}
\usepackage[nottoc,notlot,notlof]{tocbibind}
\usepackage[title]{appendix}

\usepackage{tikz}
\usetikzlibrary{positioning}
\usetikzlibrary{intersections}
\newlength{\PicScale}
\usepackage[active]{srcltx}
\usepackage{latexsym}
\usepackage{amsmath}
\usepackage{amsfonts}
\usepackage{mathrsfs}
\usepackage{dsfont}
\usepackage{amscd}
\usepackage{verbatim}
\usepackage{xcolor}
 \csname
@addtoreset\endcsname{equation}{section}

\usepackage{lipsum}
\usepackage{xargs}
\usepackage[normalem]{ulem}

\usepackage[colorinlistoftodos,prependcaption,textsize=tiny]{todonotes}
\newcommandx{\unsure}[2][1=]{\todo[linecolor=red,backgroundcolor=red!25,bordercolor=red,#1]{#2}}
\newcommandx{\change}[2][1=]{\todo[linecolor=blue,backgroundcolor=blue!25,bordercolor=blue,#1]{#2}}
\newcommandx{\Snote}[1]{\todo[backgroundcolor=red!25,bordercolor=red,noline]{S:#1}}
\newcommandx{\Anote}[1]{\todo[backgroundcolor=blue!25,bordercolor=blue,noline]{A:#1}}

\DeclareMathAlphabet{\mathscr}{OT1}{pzc}{m}{it}

\def\p{\partial}

\def\a{\alpha}
\def\b{\beta}

\def\d{\delta}

\def\l{\lambda}

\def\n{\nu}

\def\s{\sigma}

\newcommand{\be}{\begin{equation}}
\newcommand{\ee}{\end{equation}}
\newcommand{\ba}{\begin{aligned}}
\newcommand{\ea}{\end{aligned}}
\newcommand{\ben}{\begin{displaymath}}
\newcommand{\een}{\end{displaymath}}
\newcommand{\bea}{\begin{eqnarray}}
\newcommand{\eea}{\end{eqnarray}}

\def\tr{\text{tr}}

\def\12{\frac{1}{2}}

\def\eNm1{\overset{\scriptscriptstyle{(N-1)}}{e}}

\begin{document}

\begin{titlepage}       \vspace{10pt} \hfill

\vspace{20mm}

\begin{center}

{\large \bf  Analyticity Constraints and Trace Anomaly Matching}

\vspace{30pt}

Adam Schwimmer$^{a}~$ and Stefan Theisen$^{b}$
\\[6mm]

{\small
{\it ${}^a$ Weizmann Institute of Science, Rehovot 76100, Israel}\\[2mm]
{\it ${}^b$Max-Planck-Institut f\"ur Gravitationsphysik, Albert-Einstein-Institut,\\
14476, Golm, Germany}
}

\vspace{20pt}

\end{center}

\centerline{{\bf{Abstract}}}
Using an unambiguous characterization of Trace Anomalies a general proof of matching for Type A and B anomalies 
in the broken phases of Conformal Field Theories is given.
The general constraints on amplitudes of energy-momentum tensors and dilatons 
in the broken phase, which follow from matching, are analyzed.
\vspace*{5mm}
\noindent

\vspace{15pt}

\end{titlepage}

\tableofcontents{}
\vspace{1cm}
\bigskip\hrule

\section {Introduction}

Trace anomalies, reviewed in \cite{Duff},  differ in a basic way from other  anomalies which appear  in Quantum Field Theory.
While the latter -- in this general discussion we will call them ``chiral" anomalies -- have a clear topological
interpretation, relating the anomaly to characteristic classes of various gauge bundles, such an interpretation
does not exist for trace anomalies. This has its origin already at the  formulation of the anomalies: while
chiral anomalies appear as phases in the transformation of the Euclidean generating functional,
trace anomalies are real in Euclidean signature.

Another significant difference of trace anomalies is the nonvanishing of
type B trace anomalies\footnote{We will specify the classification into type A and type B trace anomalies
in Section 2.}  for a constant gauge parameter, a unique feature for anomalies of continuous symmetries,
first pointed out in \cite{DDI}.

Given the special features mentioned above, the only available technique to study trace anomalies
remains the local/perturbative one. In particular the full use of the analytic structure of correlators is indispensable.

In the present paper we summarize how the  use of analyticity allows the extraction of maximal information about trace anomalies.
We discuss two related aspects.
First we formulate exact unambiguous sum rules which characterize the anomalies
in terms of integrals over discontinuities of unambiguous dimension $-2$
invariant amplitudes. This structure is the same for both types of anomalies, A and B.
For type B anomalies we disentangle the relation between the explicit UV violation in
dilations and the nonvanishing of the trace anomaly
for constant Weyl gauge parameter.

Then we discuss  in detail how the above features influence one of the major
consequences of the existence of anomalies, `t Hooft anomaly matching.
The `t Hooft matching is there only  when the anomalous structure exists along the RG flow. For trace anomalies
therefore one can have a nontrivial `t Hooft matching only for the spontaneously broken phase of a 
Conformal Field Theory (CFT) where there is a nontrivial ``flow" between the UV and IR while
the algebra of conformal symmetry is preserved.

For chiral symmetries the modern version of `t Hooft's
original argument uses the  anomaly inflow based on the existence of the topological action in $d+1$ dimensions, 
reproducing the anomaly of the $d$-dimensional Quantum Field Theory.
For trace anomalies such  an  argument does not exist and we have again to use a detailed analysis based on the
analytic structure.  The result of the analysis is  that trace anomalies are matched between the unbroken and
broken phases, both for the type A and type B. We summarize the general arguments  for
matching: same analytic structure of the amplitudes and equality in the deep Euclidean limit. 
However, while in the unbroken phase the normalization of the type B anomaly is related to the normalization
of the explicit violation of dilations by the UV cut-off, this connection breaks down in the broken phase where the violation 
of dilations now also comes from the presence of the symmetry breaking scale. 

Once the matching is
established we should be able to use it to get further, universal information about the structure of the broken phase.
The new information  involves the dilaton  and the energy-momentum (e-m) tensor and their couplings to the
massive states in the broken phase. An efficient way to study this structure is to analyze dispersive sum rules of
dilatons and e-m tensors constrained by anomaly matching.
We discuss the derivation  and general structure of these classes of sum rules.

In Section 2 we review the structure of trace anomalies in $d=4$. We analyze in detail the
analytic structure using that both the type A and type B anomalies are reduced to completely
unambiguous relations between dimension $-2$ invariant amplitudes. We give expressions of
the anomaly in terms of convergent integrals of the discontinuities
of dimension $-2$ amplitudes and make precise the conditions under which the anomaly is related to the action of the
generators of the conformal group on the correlators.

In Section 3 we give a general proof of anomaly matching in the
broken phase. We discuss the structure of the flow and anomaly integrals  as a function
of the spontaneous breaking scale. For type B anomalies we show that while the anomaly is matched, 
the relation between the action of the dilation charge 
on the correlators and the anomaly is no longer valid in the  broken phase.

In Section 4 we discuss the metric and
dilaton source dependent generating functionals. Using their general properties and the
structure of anomalies discussed in Section 3, we give a  prescription for the
limit in the deep infrared -- defined as the breaking scale going to infinity limit --
for invariant amplitudes involving e-m  tensors and dilatons.
Using dispersion relations this gives sum rules for these amplitudes and represents the most
general consequences of anomaly matching, giving constraints on the massive spectrum and its couplings to the dilaton in the broken phase.

In Appendix A we give a general argument showing explicit matching of type A and B anomalies in the broken phase at
one loop level. The argument can be seen also as a simplest
saturation of the sum rules  derived in Section  4. In  Appendix  B we analyze in detail an explicit 
theory which was claimed as a counterexample for matching of type B. We verify that type B anomalies 
are matched, as required by the general argument.\footnote{We thank V. Niarchos for confirming this.} 

\section{Review of Trace Anomalies}
In this section, after a general review of trace anomalies, we will discuss two specific features which are essential in the matching of 
the anomalies in the broken phase. 
We start by recalling the standard framework of trace anomalies; cf. \cite{Duff} and references therein.
Let us consider a CFT which has a conserved and
traceless energy-momentum tensor $T_{\mu\nu}$. The vacuum correlator of any number of
e-m tensors contains the information about the trace anomalies.
Correlators of integer dimensional primaries and e-m tensors may have additional type B
anomalies.  By the standard device of gauging, the Ward identities (WIs) are translated into a gauge
invariance of the generating functional. In our case the ``gauge field" is the space-time metric
and the conservation and tracelessness  of the e-m tensors correspond  to the invariance of the
generating functional under  diffeomorphisms (diffeo) and  Weyl transformations, respectively.
We consider therefore the CFT coupled to a c-number source space-time metric
$g_{\mu\nu}$, such that the classical action is invariant under diffeo transformations
with parameters $\xi^{\mu}(x)$ and Weyl transformation with parameter $\sigma(x)$.
The transformation of the CFT fields is accompanied by the transformation of the metric:
\be\label{eq1}
\d_\xi g_{\mu\nu}=\nabla_\mu\xi_\nu+\nabla_\nu\xi_\mu
\ee
\be\label{eq2}
\d_\s g_{\mu\nu}=2\,\s\,g_{\mu\nu}
\ee
The correlators of e-m tensors are the coefficients of the expansion
of the generating functional in powers of $h^{\mu\nu}$, the deviation of the metric from the flat one:
\be\label{eq3}
h^{\mu\nu}=g^{\mu\nu}-\eta^{\mu\nu}
\ee
The transformation of $h^{\mu\nu}$ is therefore
\be\label{eq300}
\d_\s h^{\mu\nu}=-2\,\s\,\big(\eta^{\mu\nu} +h^{\mu\nu}\big)
\ee
Due to the transformations \eqref{eq1} and  \eqref{eq300} the WIs following from the conservation and  
tracelessness  of the e-m tensor relate correlators of different numbers of operators.

The anomaly problem is set up at the correlator level by studying the implementation of the WIs
corresponding to conservation and tracelessness. The discontinuities (``imaginary parts'') obey
the WIs  and the possible anomalies appear in real parts (subtraction polynomials) which cannot
be removed without violating at least one class of the WIs. We stress that the anomalies always
appear in polynomials of the external momenta while the negative dimension invariant amplitudes are unambiguous.

This scheme translates into the formalism using the generating functional: the anomalies are local
expressions in the gauge variations which cannot be removed by variations of local terms.
This cohomological structure was understood a long time ago giving two types of non-removable
variations/anomalies \cite{Bonora,DS}.
One can retain diffeo invariance and then the anomaly appears
in the Weyl variation which  can give two classes of cohomological structures in even dimensions:

a) Type A,  i.e. the Weyl variation is:
\be\label{eq5}
\d_\s W=a\int d^{2n}x\,\sqrt{g}\,\s\,E_{2n}
\ee
where $E_{2n} $  is the Euler density in $d=2\,n$ dimensions and

b) Type B where the Weyl variation is
\be\label{eq6}
\d_\s W=\sum c_i\,\int d^{2n}x\,\sqrt{g}\,\s\,I_i
\ee
where $I_i$  are Weyl invariant polynomials in the curvatures and their derivatives, with a total of
$2\,n$ derivatives. Their number depends on the
dimension, e.g. one in $d=4$, three in $d=6$.

An important distinction between the two types is that while \eqref{eq5} vanishes for constant
$\sigma$,  \eqref{eq6} does not.  We will call generically a trace anomaly to be Type B if it does not
vanish for $x$-independent Weyl parameter $\sigma$. This condition automaticaly  excludes cohomologically trivial
solutions of the Wess-Zumino consistency condition.

Trace anomalies can appear also in correlators involving additional operators. 
An example of a type B anomaly in the above sense, which  we will use in the following, 
involves dimension $2$ scalar operators coupled to dimension $2$ sources $J$  in $d=4$. Then the anomaly
is defined by a variation
\be\label{eq799}
\d_\s W=\tilde c \int d^{4}x\,\sqrt{g}\,\s\,J^2
\ee

We will now analyze the detailed structure of the trace anomalies in $d=4$,  in  particular their 
characterization in terms of unambiguous, regularization independent information which we will 
then use in the proof of anomaly matching. In $d=4$ the anomaly structure is determined by the 
three-point correlator which is the first anomalous one. We believe that the structure we
discuss is universal in $d=4$ for type A and B trace anomalies \eqref{eq5} and \eqref{eq6} 
as well as for the additional anomalies 
\eqref{eq799}. In higher (even) dimensions a detailed analysis has not yet been performed.
In \cite{ST2}  we presented a detailed kinematical analysis for the aforementioned set-up which
we will not repeat. We will however stress its general features  and refer to \cite{ST2} for explicit
checks.

a) Our starting point is the cohomological analysis which tells us what kinematical structure admits
an unremovable real part. We make explicit this structure through a decomposition of the correlators
into invariant amplitudes. The anomaly appears as  an unremovable constant in the r.h.s. of
combinations of invariant amplitudes.

b) Since the anomaly appears in the Weyl transformation, the diffeo WIs are obeyed and we can
use them to relate the invariant amplitudes entering the Weyl WI. Using this we can eliminate all the
dimensionless and positive dimension invariant amplitudes and the anomaly equation acquires
a universal form for both type A and type B
\be\label{eq7}
\sum_{i=1}^3 s_i\, E_i(s_1,s_2,s_3)=c
\ee
where $c$ represents the type A or type B anomaly normalizations.

c) In the universal anomaly equation \eqref{eq7} $s_i\equiv q_i^2$ where $q_i$ are the momenta
entering the three-point correlator. The invariant amplitudes $E_i$ have dimension $-2$ and $c$ in the
r.h.s. represents the anomaly constant. We stress the universal form of \eqref{eq7}:
it is equally valid for type A and type B anomalies. Even though there are 
dimensionless and positive dimension invariant amplitudes which contain single logarithmic
dependence, these amplitudes are eliminated using the diffeo invariance.
The dimension $-2$ amplitudes have pure power-like dependence on the invariants $s_i$, as
dictated by conformal invariance. The $a$ and $c$ anomalies of the three e-m correlators
appear in more than one grouping of three dimension $-2$ amplitudes with the same anomaly in the rhs.
This is a result of the relations imposed by the diffeo and Weyl WIs. For the special case
of the type B anomaly related to the dimension $-2$ scalar primary operator, there are only three dimension
$-2$ amplitudes and \eqref{eq7} is unique in this case. We will analyze now \eqref{eq7} using 
analyticity properties of the amplitudes to infer the general consequences of the anomaly.

Let us take the discontinuity of \eqref{eq7} in the invariant $s_j$, while holding the other two invariants fixed:
\be\label{eq8}
\sum_{i=1}^3 s_i\,{\rm Im}_j E_i=0
\ee
Here ${\rm Im}_j$ stands for $\frac{1}{2i}$ times the discontinuity in the variable $s_i$, with the
other two invariants kept fixed.
Then, since the discontinuities have a power behaviour dictated by the dimension $-2$ of the amplitude, we obtain
\be\label{eq9}
{\rm Im}_j E_j=-\sum_{i\neq j}\frac{s_i}{s_j}{\rm Im}_j E_i\sim\frac{1}{s_j^2}~~~{\rm for}~~~s_j\to\infty
\ee
It follows that  $E_j$ obeys a dispersion relation in the $s_j$ invariant in which for $s_j\to\infty$
the $s_j$ dependence can be taken outside the integral:
\be\label{eq10}
E_j(s_j,\{s_{i\neq j}\})=\frac{1}{\pi}\int_0^\infty ds\frac{{\rm Im}_j E_j(s,\{s_{i\neq j}\})}{s-s_j}
\xrightarrow[]{s_j\to\infty}
-\frac{1}{\pi}\frac{1}{s_j}\int_0^\infty ds\,{\rm Im}_j E_j(s,\{s_{i\neq j}\})
\ee
Now we can use \eqref{eq7} in the $s_j\to\infty$ limit using \eqref{eq10} and the fact that for $i \neq j$
the contributions of $E_i$ are suppressed by the ratio $\frac{s_i}{s_j}$ to obtain the sum rule
\be\label{eq11}
\int ds\,{\rm Im}_j E_j(s,\{s_{i\neq j}\})=-\pi\,c
\ee
for each dimension $-2$ invariant amplitude entering the basic anomaly equation \eqref{eq7}. 
It is valid at arbitrary fixed $s_i, i\neq j$. We can extrapolate it to $s_i=0, i\neq j$ assuming
that the point is nonsingular. We then obtain
\be\label{eq12}
{\rm Im}_j E_j(s_j,s_{i\neq j}=0)=-\pi\,c\,\delta(s_j)
\ee
which is obeyed in free CFT and is familiar in various local chiral anomalies \cite{DZ,FSBY,CG}.
Equations \eqref{eq11} represent the most general unambiguous formulation of trace anomalies using the analytical structure 
of correlators. In our arguments we started with \eqref{eq7} which assumes already the presence of the  anomaly 
in the r.h.s. as a result of the cohomological analysis, but we could have started with the unambiguous relations fulfilled by 
the discontinuities and conclude that an anomaly should be present if \eqref{eq11} is satisfied with a non-zero 
right hand side.  

We discuss now the general features of type B anomalies which again have  implications for the understanding of anomaly matching.
As defined above the anomaly being ``type B" means that it does not vanish for a constant Weyl parameter. 
A constant gauge parameter is usually associated with the ``charge" of the respective symmetry.
This is basically true also for $\sigma$, the gauge parameter of Weyl transformations, with a slight
modification: the global symmetry associated with the Weyl and diffeo gauge symmetries is the conformal group 
and the relation to the ``gauge symmetry" is less direct.  
The conformal group $SO(d,2)$ is defined as the diagonal subgroup of diffeo and Weyl transformations
which leaves the flat metric $\eta_{\mu\nu}$ invariant. In particular dilations, which correspond
to a diffeo transformation
\be\label{eq4}
\xi^\mu=\l\,x^\mu
\ee
are accompanied by a Weyl transformation with constant Weyl parameter $\sigma=-\lambda$.
Special conformal transformation have  a similar behaviour  but  dilations are simpler to use in order  to disentangle the relation.
The  metric deviation $h^{\mu\nu}$ transforms correspondingly by a simple rescaling of the coordinates $x^\rho$:
\be\label{eq200}
\delta_{\lambda}h^{\mu\nu}=\lambda\, x^{\rho}\partial_{\rho}h^{\mu\nu}
\ee
and therefore the transformation induces a rescaling of all the coordinates  or momenta in the correlators
of energy-momentum tensors. There is no mixing in the transformation between correlators of different numbers of
e-m tensors. This leads to a general relation between Ward identities with constant Weyl parameter and the corresponding
diffeo  transformation and the action of the dilation. Following the procedure outlined in \cite{ST2} (cf. also 
\cite{OP,Cappelli,Skenderis}) 
we will write a general Ward identity relating
the action of dilations on correlators of $n-1$ e-m tensors to the Weyl anomaly with constant Weyl parameter in the
correlator of $n$ e-m tensors. The Ward identity  is valid also when conformal symmetry is spontaneously
broken and can be used to understand the relation between dilations and the anomaly in the broken phase.

The dilation ``charge'' $D$ is the space integral of the time component of the  dilation current $J_{\mu}^{(D)}$, where
\be\label{eq400}
J_{\mu}^{(D)}\equiv x^{\nu}T_{\mu\nu}
\ee
The action of dilations on an operator ${\cal O}(x)$ is given by the equal time commutator $[D,{\cal O}(x)]$.
When the operator is the energy momentum tensor we have
\be\label{eq401}
[D,T_{\mu\nu}(x)]=\big(x^{\rho}\partial_{\rho}+d\big)T_{\mu\nu}(x)
\ee
in $d$ space-time dimensions.
From \eqref{eq400} it follows that the conservation of the dilation current is a consequence of the conservation and
tracelessness of the energy momentum tensor. Since diffeos are neither anomalous nor spontaneously broken,
the behaviour of the dilations will follow from the anomaly with constant parameter and/or the spontaneous breaking
of Weyl symmetry.

We now derive a general WI which is always valid, which relates the WIs for local diffeo and Weyl transformations
to the action of dilations.
We start with the WI which is satisfied by the correlation functions of  $n$ e-m tensors which
follows from the diffeo invariance of the generating functional:
\be\label {eq900}
\ba
&\p^\mu_{(x)}\langle 0|T\big(T_{\mu\nu}(x)\,\prod_{j=1}^{n-1} T_{\rho_j\s_j}(y_j)\big)|0\rangle
+\sum_{l=1}^{n-1}\Big\{\p_\nu^{(x)}\delta(x-y_l)\,\langle0|T\big(T_{\rho_l\s_l}(x)\prod_{j\neq l}T_{\rho_j\s_j}(y_j)\big)|0\rangle\\
&\qquad\qquad\qquad+\p_{\s_l}^{(x)}\delta(x-y_l)\,\langle 0|T\big(T_{\rho_l\nu}(x)\!\prod_{j\neq l}T_{\rho_j\s_j}(y_j)\big)|0\rangle
+(\rho_l\leftrightarrow\s_l)\Big\}=0
\ea
\ee
The WI for Weyl invariance for the same correlator is
\be\label{eq2000}
\ba
\eta^{\mu\nu}\langle0|T\big(T_{\mu\nu}(x)\prod_{j=1}^{n-1} T_{\rho_j\s_j}(y_j)\big)|0\rangle
+2\sum_{l=1}^{n-1}\delta(x-y_l)\langle0|T\big(\prod_{j=1}^{n-1} T_{\rho_l\s_l}(y_l)\big)|0\rangle
=\mathcal{A}(x,\{y_j\})
\ea
\ee
where $\mathcal{A}(x,\{y_j\})$ represents the local trace anomalies: products of delta-functions 
$\delta(x-y_j)$ with altogether four derivatives. They are the $n$-th functional derivatives of $\d_\s W$ 
w.r.t. to $g^{\mu\nu}(y_j)$, evaluated at $g_{\mu\nu}=\eta_{\mu\nu}$. 

In writing \eqref{eq900} and \eqref{eq2000} we assumed  that we are in a scheme where diffeos are not anomalous and
all the anomalies can be put in the Weyl WI.

We now use the definition \eqref{eq400} in order  to combine \eqref{eq900} and \eqref{eq2000}.
After integrating the combination over $x$ we obtain
\be\label{eq402}
\int d^4 x\, \partial_{(x)}^{\mu} \langle 0 |T(J_{\mu}^{(D)}(x) \prod_{j=1}^{n-1}T_{\nu_j\rho_j}(y_j))|0\rangle-
\sum_{j=1}^{n-1} \langle 0|T\big((4+y_{j}^{\mu}\partial_{\mu}^{(y_j)})\prod_{k=1}^{n-1} T_{\nu_k\rho_k}(y_k)\big)|0\rangle=\mathscr{a}(\{y_j\})
\ee
with $\mathscr{a}(\{y_j\})=\int dx\, \mathcal{A}(x,\{y_j\})$  the local anomaly integrated over $x$,
i.e. corresponding to a constant Weyl parameter.
WIs of this type are standard: they relate local transformations to the action of charges.
Here we wrote it for the specific setup of dilation transformations related to anomalous  Weyl transformations.
In particular due the integration over $x$ only type B anomalies will contribute to $\mathscr{a}(\{y_j\})$
in  the r.h.s. of \eqref{eq402}.

The first term on the l.h.s. of \eqref{eq402} is a total derivative and one might naively discard it.
However, in momentum space, this term involves a multiplication with $q_{\mu}$ in the limit
$q_{\mu}\to 0$ and therefore only vanishes if there is no singularity at $q_{\mu}=0$.
The second term is the infinitesimal action of dilations on correlators of $n-1$ e-m tensors, cf. \eqref{eq401}.
If there is neither a singularity at $q_{\mu}=0$ which, as we will see,
is related to spontaneous breaking of Weyl symmetry, nor an anomaly, then only
the second term in \eqref{eq402} remains and it gives the expected invariance of the correlator of $n-1$ energy-momentum tensors
under dilations. Therefore, in the unbroken phase we have the anomalous dilation Ward identity 
\be\label{eq4021}
-\sum_{j=1}^{n-1} \langle 0|T\big((4+y_{j}^{\mu}\partial_{\mu}^{(y_j)})\prod_{k=1}^{n-1} T_{\nu_k\rho_k}(y_k)]\big)|0\rangle=\mathscr{a}(\{y_j\})
\ee
while equation \eqref{eq402}, after moving  the first term to the r.h.s., gives the action  of dilations when
anomaly and/or spontaneous breaking are present.\footnote{It is amusing to consider how the identity \eqref{eq402} is satisfied
by correlators derived from  the Riegert action \cite{Riegert,Deser1,Deser2}. Since
this action does not have a scale, the second  term in \eqref{eq402} vanishes. Therefore the first term
cannot vanish and it implies an asymptotic fall-off for the correlators slower than the correct one, as pointed
out in \cite{OE}}

The correlators can be generalized to include conformal primaries
by coupling them to sources which transform homogeneously under Weyl transformations.
For instance for operators ${\cal O}_i$ with conformal weights $\Delta_i$, the dilation WI is
\be\label{eq403}
\int d^4 x\, \partial_{x}^{\mu} \langle 0 |T(J_{\mu}^{(D)}(x) \prod_{j=1}^{n}{\cal O}_j(y_j))|0\rangle-
\sum_{j=1}^{n} \langle 0|T\big((\Delta_j+y_{j}^{\mu}\partial_{\mu})\prod_{k=1}^n{\cal O}_j(y_k)]\big)|0\rangle=\mathscr{a}(\{y_j\})
\ee
It depends on the set of conformal dimensions $\Delta_j$ whether there is an anomaly or not.
One can give a similar treatment for the action of the special conformal transformations.
We will not discuss it here since it does not give additional information
about the structure of anomalies compared to the one above.

We return now to the relation between the explicit violation of dilation invariance by the dependence on the ultraviolet scale in
correlators and the type B Weyl anomaly for a constant Weyl parameter. To establish their relation in the unbroken phase we will
use \eqref{eq402}. We will later do a similar analysis for the broken phase.
In a CFT correlators of  integer dimension primaries and in particular e-m tensors can have a dependence on the UV scale.
For such a correlator a nonzero variation under rescaling of the coordinates (respectively momenta), i.e. a violation of dilations is possible.
The nature of the violation is however very much restricted by the conformal invariance of the theory.
In particular the imaginary parts, which are finite, should respect conformal invariance, i.e. for correlators of e-m tensors
they respect the diffeo and Weyl invariance WIs. Then, if  in the corresponding invariant amplitudes
the high momentum behaviour of the imaginary part is constant, one can have a single logarithm
with the UV cutoff $ \Lambda$ in the full amplitude. The rescaling of momenta implied by dilations will then give  a constant which multiplies
the invariant structure inherited from the imaginary part. The invariant
structures are of course the integrands of the type B Weyl anomalies and this fact is at the origin of the
connection between the two.

As a concrete example we consider the two-point function of the e-m tensor in $d=4$ for unbroken
conformal symmetry:
\be\label{TT}
\langle T_{\mu\nu}(p)\,T_{\rho\s}(-p)\rangle=\Big(\pi_{\mu\nu}\pi_{\rho\s}
-\frac{3}{2}\big(\pi_{\mu\rho}\pi_{\nu\s}+\pi_{\mu\s}\pi_{\nu\rho}\big)\Big)f(p^2)
\ee
where
\be
\pi_{\mu\nu}=p^2\,\eta_{\mu\nu}-p_\mu\,p_\nu
\ee
and
\be
f(p^2)=c\,\log(p^2/\Lambda^2)
\ee
Expression \eqref{TT} is easily checked to be traceless and conserved.
It is the first term in the chain generated by
\be\label{dily}
c\int d^4 x\, \!\sqrt{g}\,C_{\mu\nu}{}^{\rho\sigma}\log\left(\frac{\Box_0}{\Lambda^2}\right)C_{\rho\sigma}{}^{\mu\nu}
\ee
where $\Box_0$ is the ``flat" Laplacian contracted with $\eta_{\mu\nu}$ and $C_{\mu\nu}{}^{\rho\sigma}$
is the Weyl tensor. Collecting all the terms with $n$ sources $h^{\mu\nu}$ and Fourier transforming gives the kinematical
structure in the $n$ energy-momentum correlators which is anomalous under dilation:
when the momenta are rescaled one gets back the structure without the logarithm,
i.e. the integral corresponding to the Weyl anomaly for constant parameter in \eqref{eq6}.

All correlation functions of e-m tensors derived
from these terms obey the Ward identities which follow from diffeo  and Weyl invariance
(conservation and tracelessness of the e-m tensor, respectively),
but under dilations they reproduce the structures of the Weyl anomaly for a constant Weyl parameter \cite{DDI}.
The structures being the same we have to check the normalization. For this we use   \eqref{eq402}:
the first term vanishes and therefore one has  a direct relation between the explicit breaking  under dilations and the Weyl anomaly.
In particular we see  that  the two-point function violates  dilations with a normalization given by the Weyl anomaly, but it is traceless.
We stress that the relevant type B Weyl  anomaly starts with the three-point function following the pattern
\eqref{eq7}, i.e. it is completely determined  by unambiguous dimension $-2$ amplitudes. However, by \eqref{eq402} 
it is related to the explicit UV violation of dilations in the correlator of two e-m tensors.

The two specific features discussed in this section, i.e. the characterization of trace anomalies in terms of dimension $-2$ 
amplitudes and the relation between the type B anomaly and the explicit violation by the UV scale will help us to understand 
anomaly matching in the spontaneously broken phase.

\section{Trace Anomaly Matching in the Spontaneously Broken Phase}

The spontaneously broken phase is defined by  a field which transforms nontrivially under the conformal
group having a nonvanishing expectation value in the Poincar\'e invariant vacuum. This introduces a scale $v$
into the theory. In particular if a scalar primary operator ${\cal O}$ with dimension $\Delta$ has a non-zero expectation
value we have
\be\label{eq13}
\langle{\cal O}\rangle=v^\Delta
\ee
Generically the spectrum in the broken theory is massive with $v$ giving the mass scale. There are a few general
features of the broken phase which we will use in our analysis of the trace anomalies in this phase:

a) The Goldstone theorem implies the existence of a massless scalar $\varphi$, the dilaton. Intuitively this massless mode
originates in the exponent of \eqref{eq13} when one transforms  the expectation value. As a consequence the Weyl transformation
of the physical dilaton field will start with a shift in $\sigma$, the Weyl parameter.

b)  The dilaton
has a linear coupling to the  e-m tensor:
\be\label{eq14}
\langle 0|T_{\mu\nu}|\varphi(q)\rangle=-\frac{f}{3}\,q_\mu q_\nu
\ee
where the scale $f$ is proportional to $v$.

c) The WIs obeyed by correlators of the e-m tensor are identical with the ones in the unbroken phase simply
because the operator algebra is unchanged.
The essential difference with the unbroken phase is, however, that the e-m tensor can
couple either directly to the states in the spectrum or through the dilaton.
Due to the linear coupling to the e-m tensor and the masslessness
of the dilaton, this coupling produces an effective breaking of the
tracelessness Ward identity. We stress again that as long as we do not make the above separation, the
WIs look the same in the broken and unbroken phases. This is further elaborated in Appendix A.

d) The UV structure of the correlators is the same as in the unbroken phase since at short distances
(large momenta) compared with the scale $f$  (i.e. in the UV region) the correlators are unchanged.

Using the above features we conclude that the possible anomaly structure is the same in the broken phase as in the unbroken phase:
decomposing in invariant amplitudes formally the relations imposed by the WI are the same. The subtraction
structure for dispersion relations for the invariant amplitudes is the same due to the same high invariant behaviour
and therefore also the  compatibility of the subtractions,  i.e. the possible  anomaly structure should be the same.
Concretely if we make the decomposition in invariant amplitudes and  we eliminate, using diffeo invariance,
the positive and zero dimension amplitudes, the Ward identities allow   a  possibly anomalous equations \eqref{eq7}, where
now the amplitudes can depend also on the breaking scale $f$ while the anomaly coefficient
$\tilde c$ is a  dimensionless and therefore $f$-independent pure number 
\be\label{eq15}
\sum_{i=1}^3  s_i\,\tilde E_i(s_1,s_2,s_3;f)=\tilde c
\ee
We see that the required structure of the anomaly is present at each value of the scale $f$ characterising
the flow. 
As before, eq. \eqref{eq15} implies sum rules for the  imaginary parts of the invariant amplitudes $\tilde E_i$. 

To prove anomaly matching in the chiral cases one can use the anomaly
inflow argument, i.e. having a universal  invariant realization in $d+1$ dimensions, which is independent
of the flow,  implies  that the normalization of the anomaly cannot change. We do not have such an
argument for trace anomalies. We can however  consider the ratios of the universal equations for
anomalies \eqref{eq7} and \eqref{eq15}:
\be\label{15a}
\frac{\sum s_i\,\tilde E_i(s_1,s_2,s_3;f)}{\sum s_i\,E_i(s_1,s_2,s_3)}=\frac{\tilde c}{c}
\ee
If we consider the ``deep Euclidean limit'' $s_i\to\infty$ with fixed ratios $\frac{s_i}{s_j}$, the amplitudes in the
broken phase should agree with those in the unbroken one since all the invariants are much larger than the
breaking scale $f$. We conclude therefore that
\be\label{eq17}
c=\tilde c
\ee
i.e. anomaly matching.
We stress that this is valid for both type A and type B trace anomalies. 

If we have several breaking patterns a similar argument works for each pattern and the corresponding 
anomalies will be all equal with the basic one given by the unbroken phase.\footnote{The result of the above 
analysis can be summarized by  specifying that 
$T_{\mu}^{\mu}$ is an operator proportional to the unit operator, with a source-dependent but state 
independent normalization, i.e. 
$T^\mu_\mu=(a\,E_4+c\, C^2)1\!\!1$,  acting on all the phases of a given CFT. 
We thank Z. Komargodski for sharing this point of view with us.}

We discuss now the special features of the type B anomaly related to the dilation transformations.
We use again \eqref{eq402} which is valid independent of whether we are in the broken or unbroken phase.
The essential difference in the broken phase is that the contribution of the first term does not vanish:
like every other operator in the theory $J_{\mu}^{(D)}$ couples to the other energy-momentum tensor
either directly or through the dilaton. The direct coupling will vanish in the $q_{\mu}\rightarrow 0$ limit,
but the limit of the coupling through the dilaton is singular and contributes. From  the operatorial equation
\be\label{eq404}
\partial^{\mu}J_{\mu}^{(D)}=T_{\mu}^{\mu}+x^{\nu}\partial^{\mu}T_{\mu\nu}
\ee
we conclude that since the coupling through the dilaton respects the conservation of the energy momentum tensor only
the first term in \eqref{eq404} could contribute. Its contribution does not vanish and is proportional to
the coupling of the dilaton to the other energy-momentum tensors at zero momentum.
Therefore \eqref{eq402} can be rewritten in the broken phase as
\be\label{eq405}
\ba
&\qquad \sum_{j=1}^n \langle0|T\big((4+y_{j}^{\mu}\partial_{\mu})\prod_{k=1}^n T_{\nu_k\rho_k}(y_k)\big)|0\rangle =
-{\mathscr a}(\{y_j\})+f\,\langle \varphi(0)|T\big(\prod_{j=1}^{n}T_{\nu_j\rho_j}(y_j)\big)|0\rangle
\ea
\ee
where $\langle\varphi(0)|$ is the on-shell  dilaton state  at zero four-momentum, with a similar rewriting of \eqref{eq403}. 
The interpretation of \eqref{eq405} is simple.
In the broken phase rescaling the coordinates (or momenta) in the correlator of $n$ energy-momentum tensors
is not an invariance. First, the UV behaviour of the correlator is the same as in the unbroken phase and rescaling
the arguments of the logarithms, with the UV cutoff as the scale, gives the anomaly term with the same
coefficient as in the unbroken phase. This is the origin of the first term in \eqref{eq405}.
In the broken phase dilations are spontaneously broken and
there is a new scale $f$ and a new breaking, the second term. The breaking of the conformal charges
in any correlator reduces to the coupling of the dilaton to the other operators in the correlator 
and the second term expresses this general pattern.
We remark that the second term is nonlocal  and not invariant under dilations for all the kinematical structures
while the first term is local and selects the overall Weyl invariant structure characteristic of type B.

We summarize now the matching including the relation to dilations in the broken phase:

a) The anomalies of both types match in the two phases. The anomalies are defined either by the coefficients of the 
cohomologically nontrivial Weyl variations of the generating functional, or, equivalently, 
by the r.h.s. of the sum rules \eqref{eq7} and \eqref{eq15}, 
which involve only unambiguous, dimension $-2$ invariant amplitudes.  

b) The UV behaviour of all amplitudes is the same in the two phases and the coefficient and tensor
structure of the logarithm of the cut-off scale in the anomalous correlators is the same as the coefficient
of the type B anomaly which, in turn, is the same in the two phases. 

c) Comparing \eqref{eq4021}  and \eqref{eq405} shows that  
the relation between the transformation of correlators of energy-momentum tensors under dilations and 
the type B anomaly is different in the two phases. 
There is an additional scale $f$ in the broken phase and the last, model dependent term in 
\eqref{eq405} is due to the dilaton coupling. 
While in both phases the coefficient of the explicit violation 
due to the UV scale is related to the anomaly, this term has no relation to the anomaly.

The anomaly equation \eqref{eq15} and the sum rules which follow from it 
depend on the scale $f$. Generically for finite $f$ the analytic structure
is complicated as all massive states with the mass scale given by $f$ contribute. The dilaton has of course
special contributions, but it is difficult to separate  them from the massive contributions in a model independent way.
Even at special kinematical point we do not expect any specific
signature for the dilaton. In order to single out the dilaton contributions we have to go to the ``deep IR'' limit
defined as $f\to\infty$, which we will discuss in detail below. In this limit the massive states
get infinite mass and the only dynamical degree left is the dilaton itself. The dynamics is described
by the dilaton effective action. We start by listing some of its general properties:

a) The masslessness of the dilaton following from Goldstone's theorem should be stable.
Therefore one cannot have a potential for the dilaton.

b) The action should  be IR free and  therefore the dilaton could have only derivative self-couplings
which are necessarily irrelevant.

c) The WI should be reproduced also in the ``deep IR'' limit. In particular the anomaly equations \eqref{eq15}
should be reproduced just by the dilaton effective action together with the couplings of the dilaton to the
e-m tensor.

An interesting question is the general structure of the deep IR theory,
in particular the role of additional massless fields besides the dilaton.
Constraints on their couplings to the dilaton and the e-m tensor were discussed in \cite{Luty,Penedones}.
We believe that when the correlators of e-m tensors are studied, the only consistent structure in leading order in the $\frac{1}{f}$
expansion is a factorized one, consisting of  a ``normal" CFT where the e-m tensor correlators are obtained through intermediate
massless states which are not coupled to the dilaton in leading order,
and the dilaton sector where the e-m tensors are coupled through linear couplings to the dilaton and a single dilaton loop.
Then the anomalies are matched as the sum of the above two contributions.
We will assume the validity of this structure in the following.

The general analysis of the deep IR limit will be made, in a slightly different order, in the next section.
As a result of this analysis we will get constraints on classes of dilaton -- e-m tensor
correlators also at finite scale $f$.

\section{General Structure of Dilaton--Energy-Momentum Correlators }

We consider the broken phase of a CFT and among the dynamical fields we single out the dilaton $\varphi$.
We will be interested in the generating functional $\Gamma[\varphi_c,g]$ of one-particle irreducible (1PI) diagrams with external
dilatons and energy-momentum tensors. Here $\varphi_c$ is a classical source field, which is related
to the source for the dilaton via Legendre transformation. The metric is the classical source for the
energy-momentum tensor.  We will refer to $\Gamma$ also as the effective action.
It is obtained by integrating out all dynamical fields in the broken phase, i.e. the massive fields and the
dilaton. Functional derivatives
of $\Gamma$ w.r.t.  $\varphi_c$ give the contribution from all diagrams which are 1PI w.r.t. the dilaton.
We will make the  stronger assumption that $\Gamma$ cannot be factorized  through purely
dilaton intermediate states in leading order in $f$ even beyond the one dilaton. If this was not the case
the limit $f\to\infty$ would  produce an unstable dilaton theory in the deep infrared.

We stress that in principle 
$\Gamma$ is completely calculable given the original CFT and the breaking pattern. In particular
there are no UV divergences, besides single logs, inside $\Gamma$ and no new renormalization constants. 
As a consequence all correlators derived from $\Gamma$ obey well defined  dispersion relations  
as we will discuss in detail later. 

The basic information we are going to use is that by the Goldstone theorem at any value of the breaking
scale $f$ the massless dilaton is present with a linear coupling to the e-m  tensor.
Therefore  in the deep IR limit the theory one is left with is a  massless free scalar particle 
with the special linear coupling to the e-m, the dilaton.

In $\Gamma$ we single out the terms linear and quadratic in $\varphi_c$;  they contain the Goldstone theorem
information and are present all along the flow, giving the leading term in the $f\to\infty$ limit.
We first give its general structure and then discuss the role of the different terms.  One basic requirement is,
of course, diffeo  invariance, which is obeyed by each term:
\be\label{eq18}
\Gamma[\varphi_c,g]=\frac{1}{2}\int d^4 x\,\sqrt{g}\Big(g^{\mu\nu}\,\p_\mu\varphi_c\,\p_\nu\varphi_c-\tfrac{1}{3}\varphi_c\,f\,R+\tfrac{1}{6}\varphi_c^2\,R
+\tfrac{1}{6} f^2\,R\Big)
+\Gamma_1[\varphi_c,g]+\Gamma_0[g]
\ee
The normalized propagator of the massless dilaton in flat space fixes the first term.
The second term, linear in $\varphi_c$, is required by the term in the e-m tensor
which is linear in the dilaton, such that it reproduces
\be\label{eq19}
\langle 0|T_{\mu\nu}|\varphi(q)\rangle=-\frac{f}{3}q_\mu q_\nu
\ee
in accord with Goldstone's theorem.
It can be verified by computing the e-m tensor $T_{\mu\nu}=\frac{2}{\sqrt{g}}\frac{\delta}{\delta g^{\mu\nu}}\Gamma$
in flat space ($g_{\mu\nu}=\eta_{\mu\nu}$) at linear order in $\varphi_c$.
Dealing with a CFT, the effective action is Weyl invariant, up to the anomaly which will be discussed below,
however the  Weyl transformation of $\varphi_c$ is not a priori fixed. We know that the dilaton $\tau$ with the inhomogeneous
Weyl transformation
\be\label{eq22}
\tau\to\tau+\s
\ee
originates as the exponent of the vev, but the physical massless
dilaton field coincides with $\tau$ only at the linear level. This is the case for the change of variables
\be\label{eq21}
\varphi_c=f\big(1-e^{-\tau}\big)
\ee
The Weyl-transformation of the two sources in $\Gamma[\varphi_c,g]$ is thus
\be\label{eq20}
g_{\mu\nu}\to e^{2\,\s}g_{\mu\nu}\,,\qquad \varphi_c\to e^{-\s}\varphi_c+f\big(1-e^{-\s}\big)
\ee
and it can be checked that the terms in \eqref{eq18} in the bracket transform homogeneously under Weyl
transformation, compensating the transformation of the measure. Note that $\varphi_c-f$ transforms
homogeneously, and expressed through this combination, the first four terms in \eqref{eq18}
is the action of a conformally coupled scalar field $\varphi_c-f$ of Weyl weight one.

We could remove the $\varphi_c^2 R$ term in \eqref{eq18} by a field redefinition
\be
\varphi_c=\tilde \varphi_c+{\cal O}(\tilde \varphi_c^2)
\ee
where the higher order terms must be scalars under diffeos,
but $\tilde \varphi_c$ would have a complicated Weyl transformation.
Similarly the last, $\varphi_c$-independent term in \eqref{eq19}, can be removed
modifying the Weyl transformation. In summary the terms in the generating functional,
up to the quadratic term in $\varphi_c$, are fixed by diffeo and Weyl invariance
provided that we choose a correlated  Weyl transformation of $\varphi_c$ \eqref{eq20}.

The subsequent terms in \eqref{eq18} represent dilaton -- e-m tensor interaction
terms $\Gamma_1[\varphi_c,g]$,  and the pure e-m tensor correlator terms $\Gamma_0[g]$.
Due to the anomaly matching proved in the previous section, $\Gamma[\varphi_c,g]$ obeys the (anomalous) WI
with a normalization given by the unbroken phase,
\be\label{eq23}
\frac{\d\Gamma}{\d\s}=\frac{\d\Gamma_1}{\d\s}+\frac{\d \Gamma_0}{\d\s}={\cal A}(g)
\ee
where the Weyl variation includes both sources $\varphi_c$ and $g_{\mu\nu}$
\be\label{eq24}
\frac{\d}{\d\s}=2\, g_{\mu\nu}\frac{\d}{\d g_{\mu\nu}}+\frac{\d \varphi_c}{\d\s}\frac{\d}{\d \varphi_c}
\ee
If we want to make contact with the formalism of the previous section, where we had just correlators of
e-m tensors sourced by $g_{\mu\nu}$, we have to `integrate out' $\varphi_c$. This introduces
one-dilaton reducible diagrams. This is achieved by evaluating $\Gamma$ on-shell w.r.t. the dilaton, i.e.
solve the equation
\be\label{eq25}
\frac{\delta \Gamma[\varphi_c,g]}{\delta\varphi_c(x)}=0
\ee
for $\varphi_c(x)$, which becomes a functional of $g_{\mu\nu}$,
\be\label{eq26}
\varphi_c(x)=\psi\big(g(x)\big)
\ee
Then
\be\label{eq27}
\Gamma\big[\psi(g),g\big]=W[g]
\ee
is the generating functional of connected correlation functions of e-m  tensors.
The anomaly equation becomes
\be\label{eq28}
2\, g_{\mu\nu}\frac{\delta W[g]}{\delta g_{\mu\nu}}={\cal A}(g)
\ee
i.e. it obeys the same anomalous Weyl WI  as in the unbroken phase.
Then at generic $f$, besides the contributions  in which the e-m tensor couples directly to
the massive states, there are diagrams in which it couples through the dilaton. For general $f$
the various terms in $\Gamma_1[\varphi_c,g]$ represent ``form factors" for the dilaton couplings.

The ``interaction term" in the effective action $\Gamma_1$ contains, besides terms starting with three $\varphi_c$,  also terms
with one and two $\varphi_c$ and we will make now precise how we distinguish these terms from the contributions singled out in
the first term (the ``kinetic term") in \eqref{eq18}. Let us consider the coefficient of two $\varphi_c$ in Fourier space after removing the
$\delta$-function which imposes energy-momentum conservation, $\Pi(q^2,f)$.  
Since we assumed that in the deep IR limit we have just  a free massless scalar,
this means that the contribution of $\Pi$ should be subdominant, i.e.
\be\label{eq100}
\Pi(q^2,f) \to ct\, \frac {(q^2)^2}{f^2} +{\cal O}\textstyle{\big(\frac{1}{f^4}\big)}\qquad\hbox{for}\qquad  f\to\infty
\ee
Similarly we can consider the off-shell one dilaton -- one e-m  correlator in Fourier space in $\Gamma_1$:
\be\label{eq101}
V(q^2,v) \big(q^{\mu}q^{\nu}-\eta^{\mu\nu}q^2\big)h_{\mu\nu}(q)\varphi_c(-q)
\ee
The term with this kinematical structure in the kinetic term defines the normalization of $f$
all along the flow including the deep IR. Therefore $f$ should be subdominant, i.e.
\be\label{eq102}
V(q^2,f) \to -\frac {f}{3\,q^2} +{\cal O}\textstyle{\big(\frac{1}{f}\big)}\qquad\hbox{for}\qquad  f\to\infty
\ee
We could use similar arguments to distinguish  the other terms in $\Gamma_1$ with one or
two $\varphi_c$ and any number of $h_{\mu\nu}$ from  the corresponding terms  in  the kinetic term.

A  general structure, which is independent of a specific theory, will  appear  only in the $f\to \infty$ limit,  which
we will now discuss in detail. In this limit the masses of all massive states go to $\infty$ and decouple.
The limit is survived only by the massless states. Here we distinguish between the set of massless states
which belong to an unbroken subsector of the theory, which can be factored out, and the dilaton.
Therefore the dilaton gives the full particle  content in this limit. As a consequence in this limit nonlocal terms
in the generating functional could be contributed only by dilaton loops. Such terms will have strength of
increasing positive powers of $f$ and this structure would signal the inconsistency of the deep IR limit.
Therefore the only acceptable terms are local. These terms have the maximal
positive power of $f$ and represent the real parts contributed  by the decoupled massive states.
The only exception to this general structure  will be the one loop dilaton contribution to $\Gamma_0$.

We discuss now the structure of the local terms in the $f\to\infty$ limit.
It is convenient to use the dimensionless dilaton source $\tau$ defined in \eqref{eq21}.
Then the local terms are characterized by the number $n$ of derivatives (on the metric $g$
and the source $\tau$ together) which determines the power $f^{4-n}$ normalising them.
An essential constraint is obtained from the diffeo and Weyl invariance of the generating functional.
As we discussed before, the Ward identities are obeyed once we include the dilaton and its linear
coupling to the e-m   tensor. Due to anomaly matching the
Weyl invariance is anomalous with the normalization  given by the anomalies in the unbroken phase.
The anomaly should be reproduced for all $f$ and since it is dimensionless the term responsible
in the $\tau,\,g_{\mu\nu}$ basis will have four derivatives. This term, which is completely fixed
by the anomalous symmetry transformations, is the Wess-Zumino term\be\label{29}
\ba
&-(a-1)\int d^4 x\,\sqrt{g}\Big(\tau\,E_4+4\big(R^{\mu\nu}-\tfrac{1}{2}R\, g^{\mu\nu}\big)\p_\mu\tau\,\p_\nu\tau
-4(\p\tau)^2\,\square\tau+2\,(\p\tau)^4\Big)\\
&+(c-1)\int d^4 x\,\sqrt{g}\,\tau\,C^{\mu\nu\rho\s}C_{\mu\nu\rho\s}
\ea
\ee
The above terms are normalized such as to take into account the contribution of the dilaton as a scalar 
massless particle to the trace anomalies through a single loop contained in $\Gamma_0(g)$.
Once the anomaly is matched all the other terms in the expansion should be
diffeo and Weyl invariant, except for possible local counter terms which give rise
to cohomologically trivial anomalies,  which will be discussed below. 

The possible Weyl invariant local terms can be easily enumerated.
Define the Weyl invariant combination
\be\label{30}
\hat g_{\mu\nu}=e^{-2\,\tau}g_{\mu\nu}
\ee
Then any diffeo invariant expression of $\hat g_{\mu\nu}$ is also Weyl invariant.
We start with the local expression with no derivatives:
\be\label{eq31}
L_0\equiv f^4\int d^4 x\,\sqrt{\hat g}=f^4\int d^4 x\sqrt{g\,}e^{-4\,\tau}
\ee
Such a term in flat background, after expressing $\tau$ in terms of the physical
dilaton field $\varphi_c$, produces a potential for the dilaton and
by the stability argument for the deep IR theory, it cannot appear.
But we could have a term with two derivatives:
\be\label{eq32}
L_2\equiv f^2\int d^4 x\sqrt{\hat g}\,\hat R=f^2\int d^4 x\sqrt{g}\,e^{-2\tau}\Big(R+6\, \p^\mu\tau\,\p_\mu\tau\Big)
\ee
where $\hat{R}$ is the scalar curvature corresponding to $\hat g_{\mu\nu}$.
We included in the effective action \eqref{eq18} the quadratic term with a normalization that
defined the linear coupling of the dilaton to the e-m tensor. After the  change of variable
\eqref{eq21} $L_2$ is identical, up to the overall normalization, with the kinetic term in \eqref{eq18}.
Thus $L_2$ is already present in the generating functional with a fixed normalization and therefore 
we will  not include  this term further.
With four derivatives we could have two independent terms after taking into account that
$\int d^4x \sqrt{\hat g}\,\hat E_4$, where $\hat E_4$ is the Euler density for $\hat g_{\mu\nu}$,  is a total derivative:
\be\label{3133}
L_4^{(1)}\equiv\int d^4 x\,\sqrt{\hat g}\,\hat R^2=\int d^4 x\,\sqrt{g}\Big(R+6\,\square\tau-6\, \p^\mu\tau\,\p_\mu\tau\Big)^2
\ee
and
\be\label{eq34}
L_4^{(2)}\equiv\int d^4 x\,\sqrt{\hat g}\,\hat C^2=\int d^4 x\,\sqrt{g}\,C^2
\ee
where $\hat{C}$  is the Weyl tensor corresponding to $\hat g_{\mu\nu}$.
Due to the Weyl transformation properties of $\hat{C}$, \eqref {eq34} is independent of the dilaton source $\tau$ (or $\varphi_c$).
Beginning with the terms with six derivatives the normalizations will involve negative powers of $f$.
Using \eqref{eq21} the invariant terms and the Wess-Zumino term can be expanded in powers of $\varphi_c/f$
and the contributions corresponding to a given number $n$ of dilatons is identified as the coefficient of $\varphi_c^n$.
We have therefore a systematic expansion of the  generating
functional $\Gamma_0+\Gamma_1$ in the $f\to\infty$ limit:
\begin{subequations}
\be\label{eq35}
\ba
\Gamma_1[\varphi_c,g_{\mu\nu},f]&\xrightarrow{f\to\infty}
b_2\,\int d^4 x\,\sqrt{\hat g}\,\hat R^2\\
&-(a-1)\int d^4 x\,\sqrt{g}\Big(\tau\,E_4+4\big(R^{\mu\nu}-\tfrac{1}{2}R\, g^{\mu\nu}\big)\p_\mu\tau\,\p_\nu\tau
-4(\p\tau)^2\,\square\tau+2\,(\p\tau)^4\Big)\\
&+(c-1)\int d^4 x\,\sqrt{g}\,\tau\,C^{\mu\nu\rho\s}C_{\mu\nu\rho\s}+{\cal O}\big(\tfrac{1}{f}\big)
\ea
\ee
\be\label{eq36}
\Gamma_0[g_{\mu\nu},f]\xrightarrow{f\to\infty}
\Gamma_0^{\hbox{\tiny 1-loop}}[g_{\mu\nu}]~~~~~~~~~~~~~~~~~~~~~~~~~~~~~~~~~~~~~~~~~~~~~~~~~~~~
\ee
\end{subequations}
where $\Gamma_0^{\hbox{\tiny 1-loop}}$ is non-local and contains the anomaly of one massless scalar, the dilaton, and the 
logarithmically divergent type B counterterm which we have discussed in Section 2. 
In eq.\eqref{eq35} we have to replace $\tau$ by the source $\varphi_c$ using \eqref{eq21}.
Since Weyl invariance is anomalous, $\Gamma_1$ is defined modulo local diffeo invariant functionals of $g_{\mu\nu}$  
which do not contain dimensionful parameters, i.e. we can add to it a term $\int d^4x\sqrt{g}\, R^2$ with arbitrary 
coefficient.\footnote{Together with the first term in \eqref{eq35} this then includes the Wess-Zumino term for the 
cohomologically trivial anomaly $\int d^4 x\,\sqrt{g}\,\s\,\Box R$, which is $-\frac{1}{12}\int d^4 x\big(\sqrt{g}R^2-\sqrt{\hat g} \hat R^2\big)$.}
This controlled asymptotic expansion is the most general, universal, model independent  information on the dilaton -- e-m  tensor
amplitudes one can extract from the diffeomorphism and (anomalous) Weyl symmetries.
It gives us comprehensive constraints
on the dilaton couplings to the massive states in the broken phase which determine, after integrating them out, the
couplings to the e-m tensor.

The asymptotic expansion  \eqref{eq35}  can be translated
into sum rules by combining it with standard analyticity
assumptions for the invariant amplitudes.
To this end we decompose \eqref{eq35}  first into correlators with a fixed number of
dilatons and e-m tensors and then each correlator into invariant amplitudes multiplying the
various Lorentz structures. Such amplitudes are real analytic functions on the first Riemann sheet for one of the
Lorentz  invariants (while keeping the others fixed)  and therefore one can write
a dispersion relation for this amplitude in the selected invariant.
Since there are no other scales, the asymptotic expansion in $f$ can be alternatively viewed 
as an expansion around zero kinematical invariants. This 
gives ``low energy theorems'' for the dependence on kinematical invariants  
which is restricted by the dimension of the invariant amplitudes.
Therefore the real parts of the amplitudes are known for a certain power of the
kinematical invariants. Since they are fixed by dispersion relations 
this will give a sum rules for the imaginary part. These sum rules are really not independent:
the diffeo and Weyl Ward identities relate invariant amplitudes for correlators of different numbers of dilatons
and energy-momentum tensors. The Ward identities were already fully used to derive the various terms
in the Wess-Zumino expression; therefore when we get contributions normalized by $a$ or  $c$ in
sum rules for different amplitudes, this is a sign that these amplitudes were related by Ward identities constraining $\Gamma$.

Another general feature is related to the convergence of the dispersive sum rules we derive.
As we mentioned above, in the broken phase there are no UV divergences and all the dispersion relations are convergent.
However the way this is realized for positive or zero dimensional invariant amplitudes is
through subtractions which are calculable knowing $\Gamma$. Once the subtractions are made the
above argument guarantees that the integrand inside the subtracted dispersion relation is physical and well defined.
We will simply make these subtractions assuming that they are known and then manipulate
the dispersion integral until its convergence is obvious dimensionally.

We start with the basic sum rules which involve the three-point functions of one dilaton and two e-m tensors.
They are obtained by the decomposition of the different terms in the 
generating functional to ${\cal O}(\varphi_c, h_{\mu\nu}^2)$ and are collected in Table 1.\footnote{Here we define the 
energy-momentum tensor as the functional derivative of the generating functional w.r.t. $h_{\mu\nu}=g_{\mu\nu}-\eta_{\mu\nu}$. 
This differs from the convention used in Section 2, and is used here mainly for comparison with ref.\cite{Penedones}. In this convention 
the Ward identities \eqref{eq1000} and \eqref{eq2000} would look different, while \eqref{eq402} is not affected.}  
If $p_{1,2}$ are the momenta carried by the two e-m tensors  and $ q$ the momentum
carried by the dilaton, 
we will discuss the  structure at the special kinematical point $p_1^2=p_2^2=0$ when the only invariant
left is $q^2$. 
The kinematical structures listed in Table 1 are not independent; they are constrained by the diffeo and
(anomalous) Weyl Ward identities, 
leading to three independent amplitudes. All those linear combinations such that only one, but the same, of the
last three columns in Table 1 has a nonzero entry, are related by the Ward identities. 
By expecting the table one finds that e.g. $C_5+C_6$ is related to $A_1+2\,A_2-q^2 B_4$ and 
$2\,C_5+C_6$ to $B_2+B_6$.  
The independent one dilaton -- two e-m tensors couplings can then
be defined as the values at $q^2=0$ of the $C_5+C_6$, $2\,C_5+C_6$ and $C_1$ amplitudes measured in units of $1/f^2$. 
The third is related to the cohomologically trivial trace anomaly $\square R$ and therefore does 
not have a physical significance.
These couplings appear only in the deep IR limit; at finite $f$ the full $q^2$ dependence matters for
anomaly matching as shown by the generic sum rules \eqref{eq15}.
The amplitudes $C_i$ in Table 1, after taking out an over all $\frac{1}{f}$ factor,
have dimension $-2$  and obey unsubtracted dispersion relations. Therefore one has
\be\label{eq80}
C_5(q^2=0)+C_6(q^2=0)= \frac{1}{\pi} \int ds'\frac{g(s')}{s'} = \frac{2(a-1)}{f^2}
\ee
where
\be\label{eq81}
C_5(q^2 =s'+i \epsilon)+C_6(q^2=s'+i\epsilon)-C_5(q^2 =s'-i \epsilon)-C_6(q^2=s'-i\epsilon) \equiv 2i g(s')
\ee
A similar sum rule can be derived for the $2\, C_5+C_6$ amplitude singling out the $c$-anomaly.

\begin{table}
\caption{\small{Decomposition to ${\cal O}(h^2,\varphi_c)$ into invariant amplitudes.
The 3rd, 4th and 5th columns have to be multiplied by $1/f^2$.
Notation: $(1)=h_1^\mu{}_\mu,\,$ $ (1|1|2)=p_1^\mu h_{1\mu\nu} p_2^\nu,\,$ $(12)=h_1^{\mu\nu} h_{2\mu\nu},\,$, etc.,  $q=p_1+p_2$;
$p_1$ and $p_2$ with $p_i^2=0$  are the momenta of the two `gravitons'.}}
\small{
\be\nonumber
\hspace*{-1cm}
\begin{array}{|c|c|c|c|c|c|c|c|c| }
\hline
&\hbox{tensor structure} &\sqrt{g}(\tau E_4+\dots)& \sqrt{g}\,\tau\,C^2 & \sqrt{\hat g}\hat R^2\\
\hline\hline
A_1& (12) &\phantom{-} \frac{1}{2}q^4&\phantom{-}\frac{1}{2}q^4&\phantom{-}9\, q^4 \\
A_2& (1)(2) &-\frac{1}{2}q^4 & -\frac{1}{4}q^4 &-3\, q^4\\
B_1& (1|12|2) & -2\,q^2 & -q^2 & -24\, q^2 \\
B_2& (2|12|1) & -2\,q^2 & -2\, q^2 &-12\,q^2\\
B_3& (1|12|1)+(2|21|2) & &  &-24\, q^2\\
B_4& (1) (2| 2|2)+ (2) (1|1|1) & &  &\phantom{-} 3\,q^2\\
B_5&  (1) (1|2|1)+(2) (2|1|2)&  &  &\phantom{-} 12\,q^2 \\
B_6&  (1) (1|2|2)+(2) (2|1|1)&\phantom{-} 2\, q^2 &\phantom{1}q^2  &\phantom{-}12\,q^2\\
C_{1}& (1|1|1)(1|2|1)+(2|1|2)(2|2|2)& &  & -12 \\
C_{2}&(1|1|1)(1|2|2)+(2|1|1)(2|2|2) & & & -12\\
C_{3}& (2|1|2)(2|2|1) +(1|1|2)(1|2|1) &  & & \\
C_{4}& (1|1|1)(2|2|2)& \phantom{-}2&\phantom{-}\frac{2}{3}& \\
C_{5}&(2|1|2)(1|2|1) &\phantom{-}2 &\phantom{-}2&  \\
C_{6}&(1|1|2)(2|2|1) &-4 &-2 & \\
 \hline
\end{array}
\ee}
\end{table}

\begin{table}\label{Table2}
\caption{\small{Decomposition of the WZ action expanded to ${\cal O}(h^2,\varphi_c^2)$ into invariant amplitudes.
The 3rd, 4th and 5th columns have to be multiplied by $1/f^2$, the 7th column by $1/f$.
$p_i$ and $k_i$ are the momenta of the gravitons and dilatons, respectively with $k_i^2=p_j^2=0$; 
the independent momenta are $p_1,p_2$ and 
$r=k_1-k_2$. $(1|1|2)=p_1^\mu h_{1\mu\nu}p_2^\n$, etc.   
If one multiplies each tensor structure by the entry in the 6th column and
adds them up, one obtains zero.}
}
\small{
\be\nonumber
\hspace*{-1cm}
\begin{array}{|c|c|c|c|c|c|c|c| }
\hline
&\hbox{tensor structure} &\sqrt{g}(\tau E_4+\dots)& \sqrt{g}\,\tau\,C^2 & \sqrt{\hat g}\hat R^2 & \hbox{Schouten}  & \sqrt{\hat g}\hat R  & \sqrt{\hat g}\\
\hline\hline
A_1& (12) &-\frac{1}{2}(t^2+u^2)& \phantom{-}\frac{1}{2} s^2& & - s t u &\phantom{-}3\,s & -6  \\
A_2& (1)(2) & -t u & -\frac{1}{4}s^2 &9(t^2+u^2)& s t u &-2\, s &\phantom{-} 3  \\
B_1& (1|12|2) &  - s & -s  && \frac{1}{2}(s^2+4 t u) & -4 & \\
B_2& (2|12|1) & - s & -2 s && \frac{1}{2}(s^2+4 t u)& -8& \\
B_3& (1|12|1)+(2|21|2) & 2\,s&  &&-\frac{1}{2}(t-u)^2&-4 & \\
B_4& (1) (2| 2|2)+ (2) (1|1|1) & -\frac{3}{4}s &  & -\frac{15}{2} s&\frac{1}{4}(t-u)^2 & \phantom{-}1 & \\
B_5&  (1) (1|2|1)+(2) (2|1|2)& -\frac{3}{4}s &  & \phantom{-}\frac{9}{2} s &\frac{1}{4}(t-u)^2 &-1 & \\
B_6&  (1) (1|2|2)+(2) (2|1|1)& \frac{3}{2} s & s &&-\frac{1}{2}(s^2 +4 t u)& \phantom{-}2 & \\
B_7&(1) (r|2|2) - (2)(r|1|1)  & \frac{1}{2}(t-u) &  &&-\frac{1}{2}s(t-u) & & \\
B_8& (1) (r|2|1) -(2)(r|1|2) & \frac{1}{2}(u-t) &  &-9(t-u)&\frac{1}{2}s(t-u)& & \\
B_9& (r|12|2)-(r|21|1) &  &  &&\frac{1}{2}s(t-u)& & \\
B_{10}&(r|12|1)- (r|21|2)  & t-u &  && -\frac{s}{2}(t-u)& & \\
B_{11}&(1) (r|2|r)+ (2) (r|1|r)  & \frac{1}{4}s & & \phantom{-}\frac{9}{2} s&\frac{1}{4} s^2& -3 & \\
B_{12}& (r|12|r) & -s &  && -\frac{1}{2}s^2&\phantom{-}12 & \\
C_{1}& (1|1|1)(1|2|1)+(2|1|2)(2|2|2)& &  & -15 & & & \\
C_{2}&(1|1|1)(1|2|2)+(2|1|1)(2|2|2) & \phantom{-}1& & & & & \\
C_{3}& (2|1|2)(2|2|1) +(1|1|2)(1|2|1) & -1 & & & & & \\
C_{4}& (1|1|1)(2|2|2)& \phantom{-}4&\phantom{-}\frac{2}{3} &\phantom{-}21 &-s & &\\
C_{5}&(2|1|2)(1|2|1) & \phantom{-}2& \phantom{-}2 & \phantom{-}9 &-s & &\\
C_{6}&(1|1|2)(2|2|1) &-6 &-2 & &2 s & &\\
C_{7}&(r|1|1)(1|2|1)- (2|1|2)(r|2|2) & & & & & & \\
C_{8}& (r|1|1)(1|2|2)-(1|1|2)(r|2|2)& & & &u-t & & \\
C_{9}&(r|1|2)(1|2|1)-(2|1|2)(r|2|1) & & & &t-u & & \\
C_{10}&(r|1|1)(2|2|2)-(1|1|1)(r|2|2) & & & &u-t & &\\
C_{11}& (r|1|2)(1|2|2)-(1|1|2)(r|2|1)& & & &t-u & &\\
C_{12}& (r|1|2)(2|2|2)-(1|1|1)(r|2|1)& & & & & & \\
C_{13}& (r|1|r)(1|2|1)+(2|1|2)(r|2|r)& -1 &  & \phantom{-}9& & &\\
C_{14}&(r|1|r)(2|2|2)+(1|1|1)(r|2|r) & -1 &  & -15 & & &\\
C_{15}&(r|1|r)(1|2|2)+(1|1|2)(r|2|r) & -1 & & &-s & & \\
C_{16}&(r|1|1)(r|2|1)+(r|1|2)(r|2|2) & \phantom{-}2 & & & & &\\
C_{17}&(r|1|1)(r|2|2) & & & &s & &\\
C_{18}&(r|1|2)(r|2|1) & \phantom{-}2 & & -36 &s & &\\
C_{19}& (r|1|r)(r|2|1)-(r|1|2)(r|2|r)& & & & & &\\
C_{20}& (r|1|r)(r|2|2)-(r|1|1)(r|2|r)& & & & & &\\
C_{21}&(r|1|r)(r|2|r) & &  & \phantom{-}9 & & & \\
\hline
\end{array}
\ee}
\end{table}

The simplest example for sum rules involving four fields is the one discussed in \cite{KS},
i.e. corresponding  to the term with four  $\varphi_c$ in $\Gamma_1$.

We discuss now  in detail the sum rules one can derive  for two dilaton--two e-m tensor amplitudes.
A sum rule of this class was proposed in \cite{Penedones}.
We stress that since we are interested in universal consequences of the structure of the broken phase,
we will not couple the e-m tensor to gravitons and neither add their self interactions,  but rather consider
correlators of e-m tensors generically off-shell, i.e. with their  unconstrained  Lorentz structure, however massless. 
We will comment later on the possibility of connecting correlators having a different number of e-m tensors through momentum 
dependent terms -- dilaton and ``graviton" propagators -- and deriving in this way new combinations of sum rules. 
For counting the independent Lorentz structures in the correlators  the helicity formalism \cite{Wick}
is useful, but the helicity wave functions introduce kinematical singularities which spoil the dispersive
sum rules based on analyticity. We will therefore use invariant amplitudes  which obey crossing symmetry
and have well defined analyticity properties.

In order to characterize the invariant amplitudes  we will, as we did for the three-point function discussed 
above,  simply use the definition of correlators following from the
expansion of the generating functional in terms of $h_{\mu\nu}=g_{\mu\nu}-\eta_{\mu\nu}$.
If the momenta of the two e-m tensors in the correlator are $p_1$ and $p_2$,
then the invariant amplitudes will be the coefficients of scalars which can be 
constructed by contracting $h_{\mu\nu}(p_1) h_{\rho\sigma}(p_2)$ with the three independent momenta and the 
Minkowski metric. The dilaton sources being scalars do not produce any
additional Lorentz index dependence. In this way one obtains 35  amplitudes. As they hide one Schouten type identity, 
the number of independent amplitudes is 34. We will not make explicit this identity:
since the information of the sum rules is carried by the asymptotic expansion which is given in a covariant
form and therefore obeying the Schouten identity, we will remember that only 34 sum rules are really independent.
Of course  the two e-m tensor -- two dilaton correlation function is related by diffeo and Weyl Ward identities
to other correlators, but again the relevant asymptotic information used is made already consistent with
these Ward identities and therefore the 34 independent sum rules are consistent.
In Table  2 we list the invariant amplitudes (column 1) together with the kinematical structures (column 2)
which define them in the expansion. We choose in addition to $p_1,p_2$ as a third independent
momentum,
\be\label{eq41}
r=k_1-k_2
\ee
where $k_1,k_2$ are the momenta carried by the dilaton sources.
Our choice of the kinematical invariants for this class of correlators is
\be\label{eq42}
\ba
(p_1+p_2)^2&=2\,p_1\cdot p_2=\!\!\!&s&=2\, k_1\cdot k_2=(k_1+ k_2)^2\\
(p_1+k_1)^2&=2\,p_1\cdot k_1=\!\!\!&t&=2\,p_2\cdot k_2=(p_2+k_2)^2\\
(p_1+k_2)^2&=2\, p_1\cdot k_2=\!\!\!&u&=2\,p_2\cdot k_1=(p_2+k_1)^2
\ea
\ee

Expanding the Wess-Zumino term and the local counterterm to order $\frac{1}{f^2}$, we can identify 
each invariant amplitude in the limit $f\to\infty$,
where it is a homogeneous polynomial in the kinematical invariants,  whose order reflects the 
dimension of the invariant amplitude which is determined by the dimension of the kinematical prefactor. 
We list these contributions in Table 2.
The normalization of the local counterterm 
is completely determined by expanding $\Gamma$, but it is model dependent. 
Therefore in our general treatment they represent  subtractions in the
dispersion relations which we cannot control. We get therefore  meaningful universal relations only if the local counterterm contribution to a
particular amplitude vanishes. We can either use the specific form of the contributions to choose
kinematical points where the specific terms vanish, or consider invariant amplitudes or linear combinations which are not affected 
by the counter term. Examples are $A_1$, which we will discuss in detail below, or 
$3 B_4 + 5 B_5$. Both are free of the $\sqrt{\hat g}\,\hat{R}^2 $  contribution and therefore obey a universal sum rule.

We discuss now how to use the low energy theorems  for the amplitudes of Table 2 to derive dispersive sum rules.
The amplitudes are  analytic  functions with  branch points in the kinematical invariants  and do not have  an additional scale 
besides $f$. In the spontaneously broken phase there are no ultraviolet singularities and in principle the correlators
are calculable  without introducing additional renormalization parameters (subtraction constants).
In the limit $f\to\infty$ the correlator gives therefore, through Cauchy’s theorem,  a genuine constraint on the amplitudes. 
Let us consider one of the amplitudes, which we denote  generically by $A$, as a function
of one of the kinematical invariants, which we denote generically by $t$, the other invariant 
being put to $0$. If the dimension of the amplitude is $d$, where $d=2,0,-2$,  the possible
contribution in the limit $f\to \infty$  has the form   $c\,\frac{t^{(d+2)/2}}{f^2}$. 
If we factor out this structure the remaining function $\psi(z) $ is a function of the dimensionless variable $z\equiv \frac{t}{f^2}$.
Concretely, we define $\psi(z)$ via
\be\label{eq997}
A(t)\equiv \frac{t^{\frac{d+2}{2}}}{f^2}\psi(z)
\ee
Then uniformly the low energy theorems correspond to 
\be\label{eq998}
\psi(0)=c
\ee
The function $\psi(z)$ has the same analyticity structure as $A(t)$, the point $t=0$ not producing a pole 
due to the low energy theorem. 
The ``maximal" behaviour of $\mathrm{Im}\,A$ is fixed by its dimension to $t^{d/2}$, which translates to 
\be\label{eq999}
\mathrm{Im}\,\psi(z) \sim \frac{1}{z}\qquad\hbox{for}\qquad z\to\infty
\ee
and therefore we have uniformly the convergent  sum rule:
\be\label{eq1000}
c= \frac{1}{\pi}\int_{\bar z}^{\infty}  dz’ \frac{\mathrm{Im}\,\psi(z’)}{z’}
\ee

We illustrate this general pattern for the sum rule for the amplitude $A_1$ 
(cf. Table 2), which is slightly more complicated  
due to the explicit implementation of the $t-u$ crossing symmetry.  
The amplitude has dimension $+2$ and therefore we  define the  function $\psi_{1}(z)$ of the dimensionless variable 
$z\equiv\frac{t^2}{f^4}$ such that
$A_1(t,s=0)\equiv \frac{t^2}{f^2} \psi_{1}(z)$.
Then  the low energy theorem becomes $\psi_{1}(z=0)=a-1$.
We can  write  a dispersion relation for $\psi_{1}(z)$ which has  discontinuity  only for  positive $z$ starting at  $\bar z>0$.
\be\label{eq1000}
\psi_{1}(z) =\frac{1}{\pi}\int_{\bar z}^{\infty}  dz’ \frac{\mathrm{Im}\,\psi_{1}(z’)}{z’-z}
\ee
and therefore the sum rule is 
\be\label{eq1001}
a-1 =\frac{1}{\pi}\int_{\bar z}^{\infty}  dz’ \frac{\mathrm{Im}\,\psi_{1}(z’)}{z’}
\ee
Here we assumed that the integral is convergent. 
This is the case if $\mathrm{Im}\, A$ has the maximal behaviour $\propto t$ as $t\to\infty$, 
corresponding to its dimension. This would mean that
$\psi_1$ behaves as $\frac {1}{\sqrt{z}}$, making  \eqref{eq1001} convergent.
Using \eqref{eq1001} and going back to the $A_1 $ amplitude and the $t$ variable we obtain the sum rule 
\be\label{eq60}
\frac{1}{\pi}\int_{M^2}^\infty dt'\,\frac{\mathrm{Im}\,{A_1(s=0,t')}}{t'^3}=\frac{a-1}{2\,f^2}
\ee
where $M^2$ is the lowest branch point proportional to $f^2$. 

Likewise, if we write $A_1(s,t=0)=\frac{s^2}{f^2}\psi_2(z)$ with $z\equiv\frac{s}{f^2}$, we have from the 
low-energy theorem $\psi_2(0)=\frac{1}{2}[(a-1)+(c-1)]$ and we find the sum-rule 
\be\label{eq61}
\frac{1}{\pi}\int ds'\frac{\mathrm{Im}\,A_1(s',t=0)}{s'^3}=\frac{(a-1)+(c-1)}{2\,f^2}
\ee

As we have already mentioned, generically the amplitudes' deep infrared limit contains  
contributions from the invariant local terms whose normalization, in principle calculable, is not known. 
Another way to eliminate these is to 
cancel nonuniversal contributions between 1PI contributions with
different numbers of sources. Such a combination is useful for sum rules
if the corresponding combination of Wess-Zumino terms continues to be local. This may require additional conditions
which need to be imposed on the kinematics.
A natural choice is to add one-particle reducible diagrams with cancelled propagators
to the one-particle irreducible contribution. A simple way to produce such combinations is to consider 
the connected one-particle reducible 
combinations  and then cancel the external propagators.

In order to produce the full  (one-dilaton reducible) generating functional we add the source term
\be\label{eq800}
\int d^4x \sqrt{g}\, J\, \varphi_c
\ee
to the effective action $\Gamma[\varphi_c,g_{\mu\nu}]$ 
and solve the resulting equation of motion for $\varphi_c$. We then evaluate
the generating functional, including the source term, on the solution, which results in $W[J,g]$, the generating
functional of connected correlation functions of dilatons and energy-momentum tensors.
If we are only interested in the contribution of  $\Gamma_1$ to first order in $b_2$ (cf. \eqref{eq35}, it suffices to take
for the solution $\bar\varphi_c(J,g)$ only  the first (the ``kinetic") term  in \eqref{eq18} into account.
This solution is
\be\label{eq801}
\bar\varphi_c=\frac{1}{\Box-\frac{1}{6}R}\left(J-\frac{f}{6}R\right)
\ee
Using the definitions  \eqref{eq21}  and \eqref{3133}, we obtain to ${\cal O}(J^2)$
\be \label{eq802}
L_4^{(1)}[\bar\varphi_c,g]=
\int d^4 x\,\sqrt{g}\left(R+6\frac{\Box\bar\varphi_c}{f-\bar\varphi_c}\right)^2
=\frac{36}{f^2} \int d^4x\, \sqrt{g}\,J^2 \left(1-\frac{1}{6\Box}R \right)^2+{\cal O}(J^3)
\ee
where the $\Box^{-1}$ and $R$ should be ordered as operators in multiple products.
After removing the dilaton source $J$, multiplying with the dilaton invariants  $ k_1^2,k_2^2$, 
which amounts to amputating the external dilaton propagators, and taking the limit
$k_i^2 \to 0, i=1,2$, 
we find that the contribution from the non-universal $L_4^{(1)}$  vanishes to ${\cal O}(J^2)$, 
independent of  any additional constraint on the momenta or the structure of $h_{\mu\nu}$.
(This is no longer true at higher orders in $J$.) In particular $L_4^{(1)}$ does not contribute 
to the two dilaton -- two e-m tensor one-particle reducible amplitude at ${\cal O}(1/f^2)$. 

One must now check whether the one-particle reducible amplitudes derived from 
the Wess-Zumino term in $\Gamma_1$  are local.
This is the case if the sources $h_{\mu\nu}$ are transverse and traceless.
We can therefore  write sum rules only for invariant amplitudes which are associated with 
those tensor structures in Table 2 which do not vanish after imposing tracelessness and 
transversality of $h_1$ and $h_2$. While the one-particle reducible amplitudes derived 
from $\Gamma$ in the $f\to\infty$ or, equivalently, in the low-energy limit, are then free of
poles, this is not true for the amplitudes at generic $f$. Therefore 
the dispersion relations one writes for the new invariant amplitudes have explicit pole
terms due to dilaton propagators. 
The ``universal" information used is, however, still the set of 1PI terms in the generating functional
of the broken phase and their $f\to\infty$ limit. 

Using a similar approach we may try to derive sum rules related to the one studied in \cite{Penedones}.
For this we consider more general  combinations of 1PI terms  in \eqref{eq18}.  We add
a kinetic term $\frac{1}{2\,\kappa^2}\int d^4 x \sqrt{g}\, R$ and perform a Legendre transformation on
$\varphi_c$ and on $h_{\mu\nu}$ at tree level. In other words, in addition to a source for
$\varphi_c$,  also a source for $h_{\mu\nu}$ is included and the resulting $\Gamma$ is evaluated on-shell
for $\varphi_c$ and $h_{\mu\nu}$. This leads to a generating functional
for amplitudes which are one-particle reducible w.r.t. to the dilaton \textit{and} w.r.t. the `graviton'.
The `graviton' propagators play a purely kinematical role, connecting 
1PI correlators derived from $\Gamma_1$ to vertices derived from the kinetic term in $\Gamma$. 
There are no graviton loops  though.
Then on-shell `S-matrix' elements at leading order in $\Gamma_1$
are considered. This amounts to amputating the external propagators,
i.e. multiplying by $\Pi _{i=1}^{4}q^2_{i}$
and taking the limit $q^2_{i}\to 0,\, i=1,...,4$. One gets again linear combinations of 1PI terms, 
i.e. `S-matrix elements' which are linear combinations of  terms in  \eqref{eq18}  
with momentum dependent coefficients (internal propagators), starting with the  original  1PI terms.  
As a result, for these combinations the contribution of $L_4^{(1)}$ to `S-matrix elements'
with two external dilatons vanishes, as explained above, 
and those from the  Wess-Zumino terms stay local, despite dilaton and
graviton poles from internal propagators, again if $h_{\mu\nu}$ is transverse and traceless as was verified explicitly in \cite{Penedones}
One has therefore once more sum rules for the same class of invariant amplitudes as those discussed above. 
We stress that these amplitudes are  different from the previous ones since now they involve
different combination of  the irreducible ones involving also `graviton' propagators. This means that 
new vertices, derived from $\Gamma$, which were not present in the combinations
produced by Legendre transforming only the dilaton, enter. The dispersion relations have again
explicit pole terms and the universal information is the same as before, iff no graviton propagators
are present in the 1PI terms themselves.

\section{Conclusions}

Using the detailed analytic structure of correlators  we have analyzed the WIs which lead  to type A and B trace anomalies.
The anomalous trace WI in the scheme where diffeos are not anomalous involve  only unambiguous dimension $-2$ invariant amplitudes.
The existence of the anomalies is related to a special behaviour of the imaginary parts of the dimension $-2$ amplitudes
expressed in general sum rules. These sum rules are  similar to the  ones obeyed by imaginary parts of
dimension $-2$ amplitudes appearing in chiral anomalies for continuous symmetries.
In particular at special kinematical points where  only one invariant is left the imaginary part reduces
to a $\delta$-function. 

The above structure is identical for type A and type B  anomalies. The difference in the cohomological structure
between the two types appears through the relation of type B anomalies to the explicit breaking of the
conformal charges by the ultraviolet cut-off.

In the broken phase the structure of  trace anomalies  of  both types is the same as in the unbroken phase.
Using this fact and the identical behaviour of the dimension $-2$   amplitudes in the two phases
at large kinematical invariants one has an easy proof of trace anomaly matching. This is valid for both types.
The difference between type B anomalies in the two phases is that in the broken phase  the normalization of type B
is no longer related to the violation of dilations which comes now also from the dependence on the scale $f$.

In the broken phase the contribution of the dilaton to the anomalies is essential in constraining its couplings
to the massive states and the structure of the broken phase itself. A systematic way to study these constraints
is through the study of the analytic structure of correlators of energy-momentum tensors. Trace anomaly matching gives
dispersive sum rules for the invariant amplitudes of the different correlators. This is a systematic way to get
complete ``universal"  information about the dilaton couplings.
A natural question is if the structure of the spontaneously broken phase in   the dispersive sum rules
can be generalized to massive flows. A natural guess is that the  dilaton field is replaced  in a massive flow
by the trace of the energy-momentum tensor which, of course, does not vanish in a massive flow \cite{Luty,KST,Hartman}.
Alternatively the trace of the energy-momentum tensor can be replaced through the Osborn equation \cite{Osborn} by the relevant
perturbing operators multiplied by their respective beta functions. A study of correlators involving energy-momentum 
tensors and the respective replacements of the dilatons would contribute to the understanding of massive flows.

\section*{Acknowledgement} We are grateful to Z. Komargodski for very useful discussions and comments on 
the manuscript and we thank E. Pomoni and V. Niarchos for correspondence. This work was supported in part by an Israel Science Foundation (ISF) center for excellence grant (grant number 2289/18) .

\begin{appendices}

\section{General One-Loop Argument}
The purpose of this Appendix is to prove that for one-loop diagrams there is anomaly matching diagram by diagram
for the contribution of massive states for correlators of the energy-momentum tensor with 
bilinears of the massive field, including correlators of e-m tensors themselves. The argument has two parts.

The first part gives a general relation for the on-shell coupling of the dilaton to a massive state.
Let us consider a massive state (for convenience a scalar) in the broken phase and the on-shell diagonal matrix element
of the energy-momentum tensor between two such states:
\be\label{eq90}
\ba
&\\
\langle M|T_{\mu\nu}|M\rangle~~=~~~~&\\
\noalign{\vskip-1.1cm}
&\begin{tikzpicture}[scale=0.4]
\draw[style=thick](0,0)--(1.2,1) node [shift={(0.4,.0)}] {$M$};
\draw[style=thick](0,0)--(1.2,-1) node [shift={(0.4,.0)}] {$M$};
\draw node [shift={(0.0,0,0)}] {$\bullet$};
\draw node [shift={(-1.0,.0)}] {$T_{\mu\nu}$};
\draw[style=thick](-1.4,0.05)--(0.0,0.05);
\draw[style=thick](-1.4,-0.05)--(0.0,-0.05);
\draw node [shift={(1.7,.0)}] {$+$};
\fill[black!25] (0.0,0.0) circle (.4cm);
\end{tikzpicture}
\begin{tikzpicture}[scale=0.4]
\draw[style=thick](0,0)--(1.2,1) node [shift={(0.4,.0)}] {$M$};
\draw[style=thick](0,0)--(1.2,-1) node [shift={(0.4,.0)}] {$M$};
\draw node [shift={(0.0,0,0)}] {$\bullet$};
\draw[style=dashed](0,0)--(-2.0,0) node [shift={(-0.9,.0)}] {$T_{\mu\nu}$};
\draw[style=thick](-2.0,0.05)--(-3.2,0.05);
\draw[style=thick](-2.0,-0.05)--(-3.2,-0.05);
\fill[black!25] (0.0,0.0) circle (.4cm);
\end{tikzpicture}
\ea
\ee
$T_{\mu\nu}$ carries $0$ four momentum.
As indicated, the matrix element has two contributions: from the direct coupling and from the coupling
through the dilaton.

The total trace of the energy-momentum tensor  should vanish as a consequence of the Weyl WI.
The contribution to the trace from the first part in \eqref{eq90} is
\be\label{eq91}
\langle M|T^\mu_\mu|M\rangle_1= M^2
\ee
which assumes that there is an improved energy momentum tensor such that the $M\to 0$ limit is smooth.
The contribution from the second part is
\be\label{eq92}
\langle M|T^\mu_\mu|M\rangle_2= \gamma\, f^2
\ee
We have used the linear dilaton--energy-momentum tensor coupling which, after taking the trace, cancels the dilaton pole.
Here $\gamma$ is the strength of the dilaton coupling.
The Weyl WI then leads to the universal relation
\be\label{eq93}
\gamma\, f^2+M^2=0
\ee
In particular, if after the breaking there is a dimension $+1$ massless scalar
in the broken phase, the dilaton coupling should vanish, in line with the constraints on dilaton couplings to further
massless states which we have mentioned above.

The second part of the argument deals with the presence of the anomaly when one studies correlators involving free
massive particles. There is a contribution from the explicit breaking proportional to the matrix element of $M^2\phi^2$,
but in addition there is the trace anomaly such that the limit $M\to 0$ is smooth in the correlators.
In the end the anomaly matching at one loop becomes simply the continuity 
condition with respect to the explicit breaking mass parameter.

We start with a schematic rederivation of the trace anomaly of a massless scalar field in 
the unbroken phase using Pauli-Villars regularization. Consider
the triangle diagram of the (improved) energy-momentum tensor and two bilinears of the scalar field.
The simplest situation is when the bilinears are simply $\phi^2$, but the procedure is clearly also valid when
the bilinears are e-m tensors in which case we have the standard $a$ and $c$ anomalies, etc.
The diagram is UV divergent and we use Pauli-Villars regularization, i.e. we add an identical diagram, with
relative minus sign, in which a Pauli-Villars field $\phi_{\rm PV}$  with mass $M_{\rm PV}$ is circulating.
Depending on the two other operators we may need more than one PV regulator diagram, all with the same mass scale, in
order to regulate also power divergences. The anomaly originates however in the logarithmic divergence which is
common to the diagrams, independent of the bilinear. We will ignore this complication and deal with the simplest case
when the bilinears are $\phi^2$. Then, in the presence of the $\rm PV$ regulator, we have two decoupled diagrams, their sum
being convergent. The diagrams correspond to a ``total energy momentum tensor" --
the sum of the (improved) energy-momentum tensors for $\phi$ and the Pauli-Villars field $\phi_{\rm PV}$ --
coupled to the ``total dimension $+2$ operator"  $\phi^2+\phi_{\rm PV}^2$.
As  a consequence the ``naive" Ward identity is obeyed for the sum of the diagrams, i.e. one has
the operatorial equation, valid in the correlator,
\be\label{eq94}
T^{\rm tot}_{\mu}{}^{\mu}=M_{\rm PV}^2 \phi_{\rm PV}^2
\ee
Therefore only the PV field contributes to the trace of the three-point function, which is
\be\label{eq95}
M_{\rm PV}^2 \langle \phi_{\rm PV}^2\phi_{\rm PV}^2\phi_{\rm PV}^2\rangle
\ee
This leads to a convergent  triangle diagram which has a well defined limit for $M_{\rm PV}\to \infty$.
This limit is the anomaly $\mathcal{A}$.

We now turn to the broken phase and study  the same correlator as above for a massive  field $\phi$.
The diagram has the same UV structure as in the massless case and we use again  a $\rm PV$ regulator.
The regulated energy momentum tensor satisfies the ``naive" Ward identity which includes now also the mass term of $\phi$:
\be\label{eq94}
T^{\rm tot}_{\mu}{}^{\mu}=M^2 \phi^2 +M_{\rm PV}^2 \phi_{\rm PV}^2
\ee
When we take the $M_{\rm PV}\to\infty$ limit  inside the correlator for \eqref{eq94}, the Ward identity
for the regulated trace contains the anomaly $\mathcal{A}$ from the $\rm PV$ term as before and a
convergent finite contribution which is the explicit breaking represented by $M^2\phi^2$,
correlated with two $\phi^2$ operators.

We combine now the two pieces of our argument in order   to prove anomaly matching.
Using \eqref{eq93} after moving the first term in the r.h.s. of \eqref{eq94} to the l.h.s. we obtain:
\be\label{eq95}
T^{\rm tot}_{\mu}{}^{\mu}+\gamma\, f^2  \phi^2 =M_{\rm PV}^2 \phi_{\rm PV}^2
\ee
Inside the correlator  the first term in the l.h.s. represents the ``direct coupling" of the trace of energy-momentum tensor
and the second term its coupling through the dilaton.
The manipulation for the second term is identical to the way \eqref{eq93}
was obtained: the momenta in the numerator of the dilaton coupling to the energy-momentum tensor
cancel the dilaton pole and one is left with the  $\gamma\,f^2$ coupling. Since the r.h.s. of \eqref{eq95}
is the anomaly, the equation proves anomaly matching at one loop order in the broken phase.

We will now discuss more complex set-ups at one loop, namely when the bilinear operators are not diagonal.
We start with a dimension $+2$ operator ${\cal O}\equiv \phi_1\phi_2$ of two different fields $\phi_1$ and $\phi_2$,
which in the broken phase get masses $M_1$ and $M_2$, respectively.
In the unbroken phase we calculate the anomaly in the usual way regulating the correlator of one energy-momentum tensor
and two ${\cal O}$ operators using $\rm PV$ regularization.
Since there are two UV divergent diagrams  corresponding to the energy-momentum tensors of the two fields,
the simplest for the counting is to consider two independent $\rm PV$ regulator fields $\phi_{{\rm PV}_1}$ and $\phi_{{\rm PV}_2}$
and therefore to obtain twice the anomaly from the  two
$M_{\rm PV}^2 \langle\phi_{\rm PV}^2\phi_{\rm PV}^2\phi_{\rm PV}^2\rangle$ terms
(actually, after taking into account the symmetry factors, the anomaly is twice bigger for the diagonal case,
but since we are interested in matching we will continue to use the above counting).
In the broken phase we will use the same two $\rm PV$ regulators and for the regulated total
energy-momentum tensor we get the equation:
\be\label{eq96}
T^{\rm tot}_{\mu}{}^{\mu}\equiv T_{1\,\mu}^{{\rm tot}\,\mu}+T_{2\,\mu}^{{\rm tot}\,\mu}
=M_1^2 \phi_1^2+M_2^2\phi_2^2  +M_{{\rm PV}_1}^2 \phi_{{\rm PV}_1}^2 +
M_{{\rm PV}_2}^2 \phi_{{\rm PV}_2}^2
\ee
Since essentially \eqref{eq96} represents the sum of two independent \eqref{eq94},
the proof of anomaly matching is immediate:
we move the mass terms to the l.h.s. and we use the relations $M_1^2+\gamma_1 f^2=0$ and $M_2^2 +\gamma_2 f^2$
to replace the mass terms with dilaton couplings.
The result is a modified \eqref{eq95} where in l.h.s. we have the direct couplings of the energy momentum tensors
and in the r.h.s. the anomaly contributed by two Pauli-Villars regulators.
An interesting particular case is when one of the masses, say $M_1$, stays zero in the broken phase.
Then, at least to one-loop order, $\phi_1$ belongs to
an unbroken CFT in the deep IR. In such a situation \eqref{eq96} tells us that since $\phi_1$ does
not have a dilaton coupling, its matching is trivial:
simply the  unbroken
anomaly is matched by the broken one even though the broken phase triangle diagram has 
massless and masive propagators.   

We stress the conclusion of this Appendix: at one loop order the anomaly matching is diagram by diagram without any
use of the specific structure of the broken phase like its overall spectrum, symmetries, etc.

In the class of models in which we have  a perturbative (loop) expansions  we can think about the calculation above as
a ``saturation" of the sum rules derived in Section 4, giving constraints on the massive spectrum.
Then at one loop level each field which became massive has an unconstrained mass, but its coupling  to the dilaton is constrained
by the relation \eqref{eq93}.

\section{Explicit Verification}

In this appendix we consider the model for which refs.\cite{Pomoni1,Pomoni2} claim that
anomaly matching is violated.
More specifically we consider ${\cal N}=2$ superconformal circular quivers with
gauge group $SU(k)$ at each of $N$ nodes and $N$ bi-fundamental
hypermultiplets, one between each pair of neighbouring nodes.
On the Higgs branch the superconformal symmetry is spontaneously
broken  (together with gauge symmetry breaking $SU(k)^N\to SU(k)$)
and therefore this is a good example to study the issue of anomaly
matching. These are interacting theories;  the analysis is restricted to the
lowest perturbative order which, in the case of dimension two scalar operators, amounts
to the computation of one-loop triangles.\footnote{In \cite{Pomoni2} also higher dimension
scalar operators were considered, leading to higher-loop diagrams for the lowest order
contribution to the three-point function. As  in \cite{Pomoni1} the mismatch was already found for
dimension two operators, we will limit ourselves to this simplest case.}
For the details of these models and their symmetry breaking we refer to the
references. Here we only summarize those features which enter the computation of
the three-point function.

There are $N$ complex scalar fields $\varphi^{(\a)}$, $\a=0,\dots,N-1$,
transforming in the  adjoint of the unbroken diagonal $SU(k)$ gauge
group; they originate from the ${\cal N}=2$ vector multiples\footnote{Here we only
need to mention the fields which were denoted $\hat\varphi$ in \cite{Pomoni1}. We have dropped
the hat for simplicity of notation.}
and their masses are
\be\label{masses}
m_\a^2=4\,v^2\,g^2\,\sin^2\frac{\pi\,\a}{N}=m_{N-\a}^2
\ee
$g$ is the gauge coupling constant which is chosen to be the same at each node,
and $v$ is the scale of symmetry breaking.
It arises from giving vevs to the scalars in
the hypermultiplets along a specific flat direction. Note that in the
broken phase one of the scalars, $\varphi^{(0)}$, remains massless. This is the
scalar in the vector multiplet of the unbroken diagonal $SU(k)$.

The dimension two operators which enter the anomalous three-point functions are\footnote{Our normalization
differs from the one in \cite{Pomoni1}.}
\be
{\cal O}^{(\a)}=\sum_{\a'}\tr\big(\varphi^{(\a')}\,\varphi^{(\a-\a')}\big)
\ee
They are chiral primary operators with protected dimension $\Delta=2$.
Here and below the sums are over $\a'=0,\dots,N-1$ and we cyclically identify
e.g. $\a\equiv\a+N$.

The dilaton $\s$ (useing the notation of \cite{Pomoni1,Pomoni2}) 
is identified as the trace of a certain real linear combination of the
hypermultiplet scalars. Its normalization is such that its propagator is $2\,i\,k\,N/q^2$.
Its coupling to the adjoint scalars is via the vertex
\be
-i\frac{\sqrt{2}}{k\,N\,v}\,\s\,\sum_{\b}m_\b^2\,\tr\big(\varphi^{(\b)}\bar\varphi^{(\b)}\big)
\ee

We want to study the correlator
\be
\Gamma^{(3,\a)}_{\mu\nu}\equiv\langle T_{\mu\nu}(-q)\,{\cal O}^{(\a)}(k_1)\,\bar{\cal O}^{(\a)}(k_2)\rangle
\ee
As already stated before, we evaluate it at lowest non-trivial order in perturbation theory, i.e. we work at one-loop,
which is a triangle. Like in the simple toy model of \cite{ST2}, there are two basic diagrams.
The first is the direct coupling of $T_{\mu\nu}$ to the triangle.
The second diagram, which is absent in the unbroken phase,
is the coupling of $T_{\mu\nu}$  to the triangle via dilaton exchange. For this we need the
linear coupling of the dilaton in the energy-momentum tensor:\footnote{This can be derived starting with the
expression of the energy-momentum tensor of the scalars in the hypermultiplets and using the definition
of the dilaton given in \cite{Pomoni1}.}
\be
T_{\mu\nu}\supset -\frac{1}{3}\frac{v}{\sqrt{2}}\big(\p_\mu\p_\nu-\eta_{\mu\nu}\square\big)\s
\ee
Using these results we find that the contribution of the dilaton diagram to the three-point function is
\be\label{dilaton_diagram}
\frac{1}{3\,q^2}\big(q_\mu\,q_\nu-\eta_{\mu\nu} q^2\big)\sum_\b 2\,m_\b^2 \big\langle\tr\big(\varphi^{(\b)}\bar\varphi^{(\b)}(-q)\big)\,
{\cal O}^{(\a)}(k_1)\,\bar{\cal O}^{(\a)}(k_2)\big)\big\rangle
\ee
Given the general decomposition of $\Gamma_{\mu\nu}^{(3,\a)}$ into invariant amplitudes
\be
\Gamma^{(3,\a)}_{\mu\nu}=\eta_{\mu\nu}\,A^{(\a)}+q_\mu\,q_\nu\,B^{(\a)}
+\big(q_\mu\,r_\nu+q_\nu\,r_\mu\big)C^{(\a)}+r_\mu\,r_\nu\,D^{(\a)}
\ee
we see that the only contribution of \eqref{dilaton_diagram} to the dimension $-2$ amplitudes $B,C$ and $D$ is
\be
B^{(\a)}_{\rm dil}=\frac{1}{3\,q^2}\sum_\b2\,m_\b^2\,
\big\langle \tr\big(\varphi^{(\b)}\bar\varphi^{(\b)}(-q)\big)\,{\cal O}^{(\a)}(k_1)\,\bar{\cal O}^{(\a)}(k_2)\big\rangle
\ee
For the anomalous Ward identity \cite{ST2}
\be
-3\,q^2\,B^{(\a)}-2\,q\cdot r\,C^{(\a)}+r^2\,D^{(\a)}=c
\ee
this implies that the contribution of the dilaton diagram can be alternatively viewed as replacing the
operatorial identity $T^\mu_\mu=0$ by
\be\label{trT}
T^\mu_\mu=2\sum_\b m_\b^2\,\tr\big(\varphi^{(\b)}\bar\varphi^{(\b)}\big)
\ee
as appropriate for massive scalar fields (cf. the discussion in Appendix A).

We now turn to the explicit computation.
The energy-momentum tensor of a conformally coupled complex scalar field is\footnote{For a massive scalar
there is the additional term $m^2\,\eta_{\mu\nu}\,\varphi\,\bar\varphi$, such that on-shell $T^\mu_\mu=2\,m^2\,\varphi\,\bar\varphi$,
cf. \eqref{trT}.}
\be
T_{\mu\nu}=\p_\mu\varphi\,\p_\nu\bar\varphi+\p_\mu\bar\varphi\,\p_\nu\varphi-\eta_{\mu\nu}\,\p^\rho\varphi\,\p_\rho\bar\varphi
+\frac{1}{3}\big(\eta_{\mu\nu}\square-\p_\mu\p_\nu\big)\varphi\,\bar\varphi
\ee
As we are only interested in the dimension $-2$ amplitudes $B,C,D$, the part of the energy-momentum tensor
proportional to $\eta_{\mu\nu}$ will not contribute and we only have to consider
\be\label{TBCD}
T_{\mu\nu}
\supset\sum_\b
\tr\left\{\frac{2}{3}\big(\p_\mu\varphi^{(\b)}\,\p_\nu\bar\varphi^{(\b)}
+\p_\nu\varphi^{(\b)}\,\p_\mu\bar\varphi^{(\b)}\big)
-\frac{1}{3}\big(\varphi^{(\b)}\,\p_\mu\p_\nu\bar\varphi^{(\b)}
+\bar\varphi^{(\b)}\,\p_\mu\p_\nu\varphi^{(\b)}\big)\right\}
\ee
The fields $\varphi^{(\b)}$, being linear combinations of scalars in the vector multiplet, transform
in the adjoint representation of $SU(k)$. Writing them as $\varphi=\varphi^A T_A$ and using the normalization
of the $SU(k)$ generators
$\tr(T_A T_B)=\delta_{AB}$, such that the $\varphi^A$ are canonically normalized,
one finds that all correlation functions of interest are proportional to
${\rm dim}(SU(k))=k^2-1$. In the following calculations we therefore ignore the matrix structure of the
scalar fields.

The contribution of \eqref{TBCD} to the first, i.e. the direct coupling diagram,  is
\be\label{direct}
\ba
&2\,C_{\mu\nu}(m_{\a-\b},m_\b,m_\b)+r_\mu\,C_\nu(m_{\a-\b},m_\b,m_\b)+
r_\nu\,C_\mu(m_{\a-\b},m_\b,m_\b)\\
\noalign{\vskip.2cm}
&\qquad\qquad\qquad\qquad\qquad+\left(\frac{1}{2}r_\mu\,r_\nu-\frac{1}{6}q_\mu\,q_\nu\right)C_0(m_{\a-\b},m_\b,m_\b)
\ea
\ee
where
\be
q=k_1+k_2\,,\qquad r=k_1-k_2
\ee
and\footnote{Here and in what follows
we drop an overall factor $\frac{\rm{dim}(SU(k))}{4\,\pi^2}$  to obtain a conveniently normalized anomalous Ward identity.}
\be
\ba
&C_0(m_1,m_2,m_2)=\int\frac{d^d\ell}{i\,\pi^{d/2}}\frac{1}{(\ell^2-m_1^2)((\ell+k_1)^2-m_2^2)((\ell-k_2)^2-m_2^2)}\\
&C_\mu(m_1m_2,m_2)=\int\frac{d^d\ell}{i\,\pi^{d/2}}\frac{\ell_\mu}{(\ell^2-m_1^2)((\ell+k_1)^2-m_2^2)((\ell-k_2)^2-m_2^2)}\\
&C_{\mu\nu}(m_1,m_2,m_2)=\int\frac{d^d\ell}{i\,\pi^{d/2}}\frac{\ell_\mu\ell_\nu}{(\ell^2-m_1^2)((\ell+k_1)^2-m_2^2)((\ell-k_2)^2-m_2^2)}
\ea
\ee
Following \cite{PV} (cf. \cite{EKMZ} for a review) we decompose
\be
\ba
C_{\mu}&=k_{1\mu}\,C_1+k_{2\mu}\,C_2\\
C_{\mu\nu}&=\eta_{\mu\nu}\,C_{00}+k_{1\mu}k_{1\nu}\,C_{11}+\big(k_{1\mu}\,k_{2\nu}+k_{1\nu}\,k_{2\mu}\big)C_{12}+k_{2\mu}\,k_{2\nu}\,C_{22}
\ea
\ee
and  express $C_1,\,C_2$  and the $C_{ij}$ in terms of $C_0$ and the basic scalar
loop integrals with one and two internal lines. 
We can then unambiguously extract from \eqref{direct} the contribution to amplitudes $B,C,D$ from
the direct coupling diagram, which we denote by the subscript $dir$.
If we add the contribution from the dilaton diagram we find the anomalous
Ward identities
\be\label{WIPomoni}
\sum_\b\Big(-3\,q^2\,B_{\rm dir}^{(\a,\b)}-2\,q\cdot r\,C_{\rm dir}^{(\a,\b)}+r^2\,D_{\rm dir}^{(\a,\b)}-2\, m_\b^2\,C_0^{(\a,\b)}\Big)=c^{(\a)}
\ee
where
\be
B^{(\a,\b)}=B(m_{\a-\b},m_\b,m_\b)\,,\quad\hbox{etc.}
\ee
We will not give the explicit expressions for the invariant amplitudes $B^{(\a,\b)}_{\rm dir}$, etc., as they are
rather lengthy. However in precisely the combination in which they appear in \eqref{WIPomoni}, great
simplifications are observed. Indeed, if we use the cyclicity in $\b$ and write the l.h.s. of \eqref{WIPomoni} as
\be
\frac{1}{2}\sum_{\b=0}^{N-1}\left(W^{(\a,\b)}+W^{(\a,\a-\b)}\right)
\ee
we find that each term in the sum evaluates to $\frac{1}{2}\times 2=1$, such that the total anomaly,
in the normalization chosen here, is
\be
c^{(\a)}=N
\ee
This holds for all values of the breaking scale $v$ and therefore for all masses \eqref{masses} and for all $\a=0,\dots,N-1$.
One can explicitly check this in the unbroken phase by setting all masses to zero, in which case only the
direct coupling diagram contributes.

Let us now look more closely at the claim of ref.\cite{Pomoni1}, that there is a mismatch for $\a\neq0$.\footnote{In
\cite{Pomoni1} a mismatch was also found for $\a=0$. In this case, from the first line
of \eqref{sumC0}, together with \eqref{eqA23} one finds $N-1$ for the contribution to the anomaly from the
massive triangles.
Later \cite{Pomoni2} it was realized  that the contribution
of a completely massless triangle, which is only present for $\a=0$, had been ignored. This is the contribution of a
scalar in the unbroken phase and once this is taken into account, matching in this sector was achieved.}
There the Ward identity was evaluated in the limit of vanishing external momenta where,
as one may convince oneself, only the dilaton diagram contributes. The analysis simplifies considerably and
one obtains
\be\label{sumC0}
\ba
&-2\sum_{\b=0}^{N-1} m_\b^2\, C_0\big(m_{\a-\b},m_\b,m_\b\big)\\
&\quad=-\sum_{\b=0}^{N-1}\Big\{m_\b^2\,C_0\big(m_{\a-\b},m_\b,m_\b\big)
+m_{\a-\b}^2\,C_0\big(m_{\b},m_{\a-\b},m_{\a-\b}\big)\Big\}\\
&\quad=-\Big\{2\,m_\a^2\,C_0(0,m_\a,m_\a)
+\sum_{\b\neq 0,\a}\Big(m_\b^2\,C_0\big(m_{\a-\b},m_\b,m_\b\big)
+m_{\a-\b}^2\,C_0\big(m_{\b},m_{\a-\b},m_{\a-\b}\big)\Big)\Big\}\\
&\quad=-\Big\{-2-(N-2)\Big\}=N
\ea
\ee
as before. Here we have used that for vanishing external momenta
\be\label{eqA23}
C_0(m_1,m_2,m_2)= -\frac{1}{m_2^2-m_1^2}+\frac{m_1^2}{(m_2^2-m_1^2)^2}\log\frac{m_2^2}{m_1^2}
\ee
with the special cases,
\be\label{C0mm0}
C_0(m,m,m)=-\frac{1}{2\,m^2}\,,\qquad
C_0(0,m,m)=-\frac{1}{m^2}
\ee
and
\be
m_2^2 \,C_0(m_1,m_2,m_2)+m_1^2\,C_0(m_2,m_1,m_1)=-1
\ee

\end{appendices}

\end{document}